\documentclass{emulateapj}
\usepackage{graphicx}

\slugcomment{~~}

\begin{document}

\title{A Distance Estimate to the Cygnus Loop  \\ 
       Based on the Distances to Two Stars Located Within the Remnant}

\author{Robert A.\ Fesen, Jack M. M. Neustadt, Christine S. Black}
\affil{6127 Wilder Lab, Department of Physics \& Astronomy, Dartmouth
  College, Hanover, NH 03755, USA} 

\and

\author{Dan Milisavljevic }
\affil{Harvard-Smithsonian Center for Astrophysics, Cambridge, MA 02138, USA \\
      Department of Physics and Astronomy, Purdue University, 
      525 Northwestern Avenue, West Lafayette, IN 47907, USA }

\begin{abstract}

Underlying nearly every quantitative discussion of the Cygnus Loop supernova
remnant is uncertainty about its distance.  Here we present optical images and
spectra of nebulosities around two stars whose mass-loss material appears to
have interacted with the remnant's expanding shock front and thus can be used
to estimate the Cygnus Loop's distance.  Narrow passband images reveal a small
emission-line nebula surrounding an M4 red giant near the remnant's eastern
nebula NGC 6992.  Optical spectra of the nebula show it to be shock-heated with
significantly higher electron densities than seen in the remnant's filaments.
This along with a bow-shaped morphology suggests it is likely red giant
mass-loss material shocked and accelerated by passage of the Cygnus
Loop's blast wave.  We also identify a B7 V star located along the remnant's
northwestern limb which also appears to have interacted with the remnant's
shock wave. It lies within a small arc of nebulosity in an unusually complex
region of highly curved and distorted filaments along the remnant's northern
shock front suggestive of a localized disturbance of the shock front due to the
B star's stellar winds.  Based on the assumption that these two stars lie
inside the remnant, combined with an estimated distance to a molecular cloud
situated along the remnant's western limb, we propose a distance to the Cygnus
Loop of $1.0 \pm 0.2$ kpc.  Although larger than several recent estimates of
$500 - 800$ pc, a distance $\simeq1$ kpc helps resolve difficulties with
the remnant's postshock cosmic ray and gas pressure ratio and estimated
supernova explosion energy.

\end{abstract}


\section{Introduction}

The Galactic supernova remnant G74.0-8.5, commonly known as the Cygnus Loop,
Veil Nebula or Network Nebula is widely considered to be a prototypical
middle-age remnant with an estimated age 
of $1- 2 \times 10^{4}$ yr
(\citep{Cox1972,McCray1979,Miyata1994,Levenson1998}).  
Discovered by William
Herschel in 1784, the remnant consists of a limb-brightened shell $2.8 \times
3.5$ degrees in angular size. 

The remnant is composed of several bright optical nebulosities including
NGC~6960, NGC~6974, NGC~6979, 
NGC~6992, NGC~6995, IC~1340, plus a large but
fainter nebulosity in the remnant's northwestern region known as Pickering's
Triangle. The near circular arrangement of these nebulae, referred to as the
Cygnus loop by early observers, quickly led to suspicions of it being a
supernova remnant \citep{Zwicky1940,Oort1946,Walsh1955}.

The remnant's large angular size, relative brightness across many wavelengths,
plus little foreground extinction due to its location more than
eight degrees off the Galactic plane ($E(B-V) = 0.05-0.15$; \citet{Ray1981}) 
has made it one of the best studied
Galactic supernova remnants.  Exhibiting an extensive and well resolved
filamentary structure, it has been shown to be an excellent laboratory for
investigating various shock processes including shock--cloud interactions,
X-ray, UV, and optical emissions from pre- and post-shock plasma, and
interstellar grain destruction. 

Based on analysis of X-ray, UV, optical, and radio observations and
modeling, the Cygnus Loop appears best understood as a remnant of a supernova
explosion which occurred inside an interstellar cavity created by winds off a
high-mass progenitor star
\citep{McCray1979,Ciotti1989,Hester1994,Levenson1997,Levenson1998,MT1999}.  

X-ray emission seen near the remnant's projected centre shows a metal-rich
plasma suggesting ejecta from a type II core-collapse event \citep{Miyata1998}.
In addition, X-ray emissions from certain regions along northeast and southwest
limbs have relatively high metal abundances also suggesting a high-mass,
core-collapse supernova 
\citep{Tsunemi2007,Katsuda2008,Kimura2009,Uchida2009,Fuji2011,Katsuda2011}.

\citet{Hubble1937} measured a proper motion of $0\farcs03$ yr$^{-1}$ of the
bright eastern and western nebulae away from the centre of expansion.  This
proper motion, when combined with an observed radial velocity of 115 km
s$^{-1}$ for 25 of the remnant's optical filaments, led to the first distance
estimate for the Cygnus Loop of $770$ pc \citep{Minkowski1958}.

% Table 1 %%%
\begin{deluxetable*}{lccl}
%\tabletypesize{\scriptsize}
\tablecolumns{4}
\tablecaption{Distance Estimates to the Cygnus Loop}
\tablewidth{0pt}
\tablehead{ \colhead{Reference} & \colhead{Distance} & \colhead{Range}  & \colhead{Method} }
\startdata
\citet{Minkowski1958}      & ~770 pc & \nodata       & bright optical filament velocities and proper motions  \\
\citet{Rapp1974}           & ~770 pc &$~470-1070$ pc & shock velocity from X-ray data analysis \\
\citet{SS1983}             & 1400 pc &$1000-1800$ pc & optical filament velocities and proper motions \\
\citet{Braun1986}          & ~460 pc &$~300-620$  pc & bright optical filament velocities and proper motions  \\
\citet{Hester1986}         & ~700 pc &$~500-1000$ pc & Balmer filament proper motion + assumed shock velocity \\
\citet{Shull1991}          & ~600 pc &$~300-1200$ pc & Fabry Perot scan velocities and filament proper motions \\
\citet{Blair1999}          & ~440 pc &$~340-570$ pc  & Balmer filament proper motion + modeled shock velocity \\
\citet{Blair2005}          & ~570 pc &$~460-670$ pc  & Balmer filament proper motion + modeled shock velocity \\
\citet{Blair2009}          & ~576 pc &$~510-637$ pc  & UV spectrum of a background sdOB star \\
\citet{Salvesen2009}       & ~640 pc &$~422-960$ pc  & ratio of cosmic ray to gas pressure in post-shock region \\ 
\citet{Medina2014}         & ~890 pc &$~790-1180$ pc & H$\alpha$ line widths \& proper motions \\
\citet{Raymond2015}        & ~800 pc &\nodata        & revised distance estimate of \citet{Medina2014} \\
This work                  & 1000 pc &$~800-1200$ pc & distance estimates to two stars located inside the remnant \\
\enddata
\label{tab:hist}
\end{deluxetable*}

Subsequent distance estimates have varied considerably with the full
range of possible distances taking into account measurement
uncertainties being 300 and 1800 pc (see Table 1). One of the most cited values is
$576 \pm 61$ pc made by \citet{Blair2009} who used the presence of O VI 
1032 \AA \ line absorption in the spectrum of a sdOB star lying in the direction to
Cygnus Loop's eastern NGC~6992 nebulosity. However, more recent distance
estimates have tended to favor numbers closer to Minkowski's original 770 pc
value.

Here we report on a distance measurement to the Cygnus Loop based on estimated
distances to two stars which we suspect lie inside the remnant.  These stars
have surrounding stellar mass loss nebulae which have properties suggestive of
an interaction with the remnant's expanding shock front. If these stars do in
fact lie within the remnant, then one can use spectroscopic parallax on these
stars to deduce the Cygnus Loop's true distance.

One star is a V = 11.6 magnitude red giant at $\alpha$[J2000] = $20^{\rm h}
56^{\rm m} 0.935^{\rm s}$, $\delta$[J2000] = $+31^{\rm o} 31' 29\farcs74$,
henceforth referred to as J205601.
It has a projected location near the remnant's bright eastern
nebulosity, NGC~6992, and appears centred within a small optical nebula quite
distinct in morphology from the remnant's other eastern limb emission features. 

This star and its surrounding nebulosity came to our attention via a color
image of the Cygnus Loop featured as ``Astronomy Picture of the Day'' for 1
December 2009 taken by Daniel Lopez\footnote{Instituto de Astrofisica de
Canarias} using the 2.5m Isaac Newton Telescope at Roque de los Muchachos
Observatory. The star's nebulosity is relatively faint, only weakly seen in the
broad band Digital Sky Survey (DSS) images, and not readily apparent even in narrow
interference filter images taken of the remnant \citep{Levenson1998}. 

A second star, BD+31~4224, is a V = 9.58 magnitude late B star located near the
remnant's northwestern limb ($\alpha$[J2000] = $20^{\rm h} 47^{\rm m}
51.817^{\rm s}$, $\delta$[J2000] = $+32^{\rm o} 14' 11.33\farcs74$).  It was
found during a follow-up search for additional stars projected with the
remnant's boundaries that exhibited possible evidence for CSM interaction with
the Cygnus Loop's shock.  This star lies within a small arc of nebulosity in an
unusually complex region of curved and distorted filaments along the
remnant's northwestern boundary suggestive of a localized disturbance in the
remnant's shock front.

\begin{figure*}[t]
\begin{center}
\includegraphics[scale=0.6]{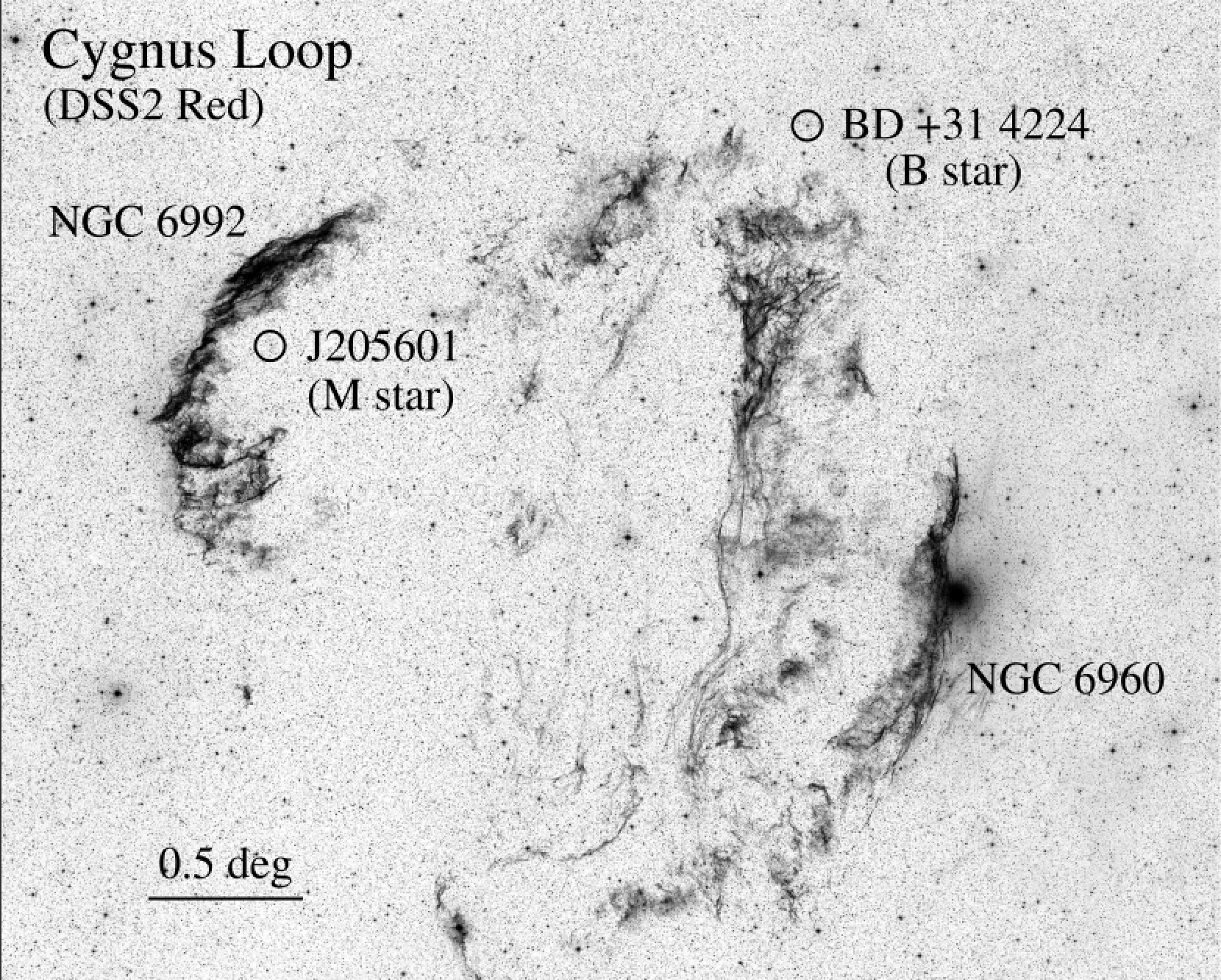}
\caption{Reproduction of the red image of the
Digital Sky Survey of the Cygnus Loop.
Marked are the two stars (J205601+313130 and BD+31 4224)
which we suspect of lying inside the remnant.
North is up, East to the left.  }
\end{center}
\label{DSS2}
\end{figure*}

The projected locations of these two stars in the Cygnus Loop 
are shown in Figure~\ref{DSS2}.  Our optical imaging and spectral observations
on both stars and the nature of surrounding nebulosities in regard to possible
physical connections to the Cygnus Loop are described in $\S$2 and
$\S$3. The stars' use as distance indicators of the remnant is discussed in
$\S$4 with our conclusions concerning the likely distance to the Cygnus Loop
summarized in $\S$5.

%%%% Figure 1

\section{Observations}

\subsection{Images}

Narrow passband images of the nebulosity around J205601 were first
obtained in  September 2011 using a 4096 x 4096 Lawrence-Berkeley National
Labs red sensitive CCD mounted on the 2.4m telescope at the MDM
Observatory at Kitt Peak Arizona.  To increase the signal-to-noise of the
images, we employed $4 \times 4$ on-chip binning which resulted in an image
scale of $0\farcs676$ per pixel. 

Images were taken using an H$\alpha$ filter (FWHM = 30 \AA), a
[\ion{S}{2}] 6716, 6731 \AA \ filter (FWHM = 50 \AA), and an [\ion{O}{3}] 5007
\AA \ filter (FWHM = 50 \AA). Three or four 600 s exposures of the nebulosity
were taken in each filter along with twilight sky flats.
Due to the size of the filters used,
the effective image field was $8.5 \times 8.5 $
arcminutes.  Standard data reduction of the images was performed using
IRAF/STSDAS\footnote{IRAF is distributed by the National Optical Astronomy
Observatories, which is operated by the Association of Universities for
Research in Astronomy, Inc.\ (AURA) under cooperative agreement with the
National Science Foundation. The Space Telescope Science Data Analysis System
(STSDAS) is distributed by the Space Telescope Science Institute.}.  This
included debiasing, flat-fielding, and cosmic ray and hot pixel removal.

Images of nebulosities around J205601 and BD+31 4224 were obtained
in June 2012 and October 2014 using the 1.3m
McGraw-Hill telescope at MDM Observatory at Kitt Peak, AZ.  These images were
taken using a 1k $\times$ 1k SITe CCD and  2$\times$2 pixel on-chip binning
yielding a image scale of 1.06 arcsec per pixel.  A series of narrow passband
[\ion{S}{2}] (FWHM $=80$ \AA) and [\ion{O}{3}] (FWHM $=50$ \AA) exposures with
exposure times of 4$\times$600 s.  

Additional narrow passband images were taken with the MDM 2.4m telescope in
September 2014 using a 2048$\times$2048 pixel SITe CCD detector with a
resolution $0\farcs508$ pixel$^{-1}$.  For J205601, two 1000 s exposures were
taken with an H$\alpha$ filter (FWHM $=80$ \AA).  For BD+31 4224,  two 600 s
exposures were taken for three nearby regions which were mosaiced into a larger
image with a field-of-view of $\approx$24$'$. Finally, photometric B and V
observations were obtained in May 2017 using the 1.3m MDM telescope and a 1k
$\times$ 1k SITe CCD. Photometric calibration observations were taken of $+50$
declination standard stars \citep{Landolt2013} observed at air masses close to
those for J205601 and BD+31~4224. 

\begin{deluxetable*}{crccccccc}
\tabletypesize{\scriptsize}
\tablecaption{Stars Suspected to Lie Inside the Cygnus Loop}
\tablewidth{0pt}
\tablecolumns{9}
\tablehead{
\colhead{Star}  &  
\multicolumn{5}{c}{\underline{~~~~~~~~~~~~~~~~~~~~~~~~~~~~~~~~~~~~~~Magnitudes\tablenotemark{a}~~~~~~~~~~~~~~~~~~~~~~~~~~~~~~~~~~~~~~}} &
\colhead{Spectral} & \colhead{M$_{\rm V}$}  & \colhead{Distance\tablenotemark{b}} \\
\colhead{ID}  & \colhead{B} & \colhead{V} & \colhead{J} & \colhead{H} & \colhead{K} &  \colhead{Type} & \colhead{(mag)} & \colhead{(kpc)}   }
\startdata
J205601\tablenotemark{c}     & $13.08 \pm 0.30$ & 11.57 $\pm 0.12$ & 7.088 $\pm 0.026$ & 6.136 $\pm 0.02$ & 5.834 $\pm 0.02$ 
                         &M4 III & $+1.0$ to $-2.0$   & 1.0 - 4.6 \\
BD+31 4224                   &  $9.53 \pm 0.02$ &  9.58 $\pm 0.02$ & 9.709  $\pm 0.02$  & 9.763  $\pm 0.02$ & 9.791 $\pm 0.02$ 
                         &B7 V-IV  & $-0.1$ to $-1.3$ & 0.8 - 1.3 \\
\enddata
\tablenotetext{a}{Magnitudes are taken from the Tycho-2 and 2MASS Catalogs.}
\tablenotetext{b}{Distances calculated assuming A$_{\rm V}$ = 0.25 }
\tablenotetext{c}{Alternate ID: TYC 2688-1037-1}
\end{deluxetable*}

\subsection{Spectra}

Low-resolution optical spectra were taken in October 2011 using the
Multi-Aperture Red Spectrometer (MARS) attached to the 4m telescope at Kitt
Peak National Observatory.  A $1.2" \times 5'$ slit was used with a 450 g/mm
VPH grating to yield a resolution of 10 \AA \ and a wavelength coverage of
$5500 - 10800$ \AA.  Spectra were obtained using east-west slits. 

Low-dispersion 4000-7400 \AA \ spectra of the filaments surrounding BD+31 4224
were obtained in October 2012 using the Boller \& Chivens CCD spectrograph at
the 2.4m Hiltner telescope at MDM Observatory. The spectrograph delivered a of
3.29 \AA\ pixel$^{-1}$ with a 1.0$''$ slit, resulting in a spectral resolution
of 12 \AA.

We also obtained spectra of BD+31~4224 and J205601 in October 2015 using the
2.4m Hiltner telescope  using the OSMOS Spectrograph \citep{Martini2011} and a blue
Grism with resolutions of 0.7 \AA~pixel$^{-1}$. The spectra covered 
3900--6800 \AA \ with a spectral resolution of 1.2 \AA \
pixel$^{-1}$. 

We performed standard pipeline data reduction using IRAF. The images were
bias-subtracted, flat-field corrected using twilight sky flats, and averaged to
remove cosmic rays and improve signal-to-noise. Spectra were similarly reduced
using IRAF and the software L.A. Cosmic to remove cosmic rays.  The spectra
were calibrated using quartz and Ar lamps and spectroscopic standard stars
\citep{Oke74,Massey90}. Relative line strengths are believed accurate to better
than 10\% for the stronger emission lines.

% Figure 2
\begin{figure*}[t]
\begin{center}
\includegraphics[width=0.85\linewidth]{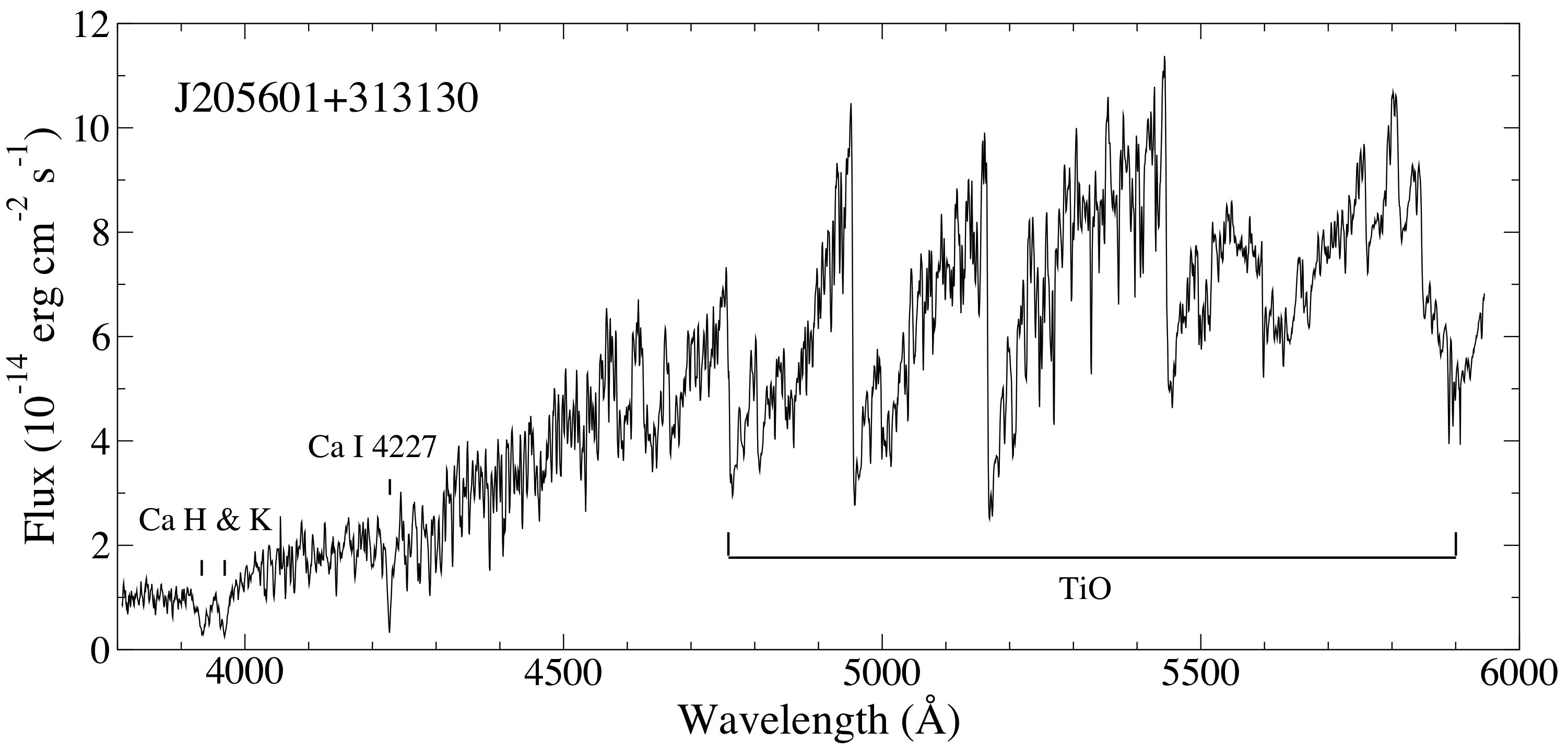}
\caption{Low dispersion optical spectra of the M star J205601+313130.  }
\end{center}
\label{M_spectrum}
\end{figure*}

% Figure 3
\begin{figure*}[t]
\includegraphics[width=1.0\linewidth]{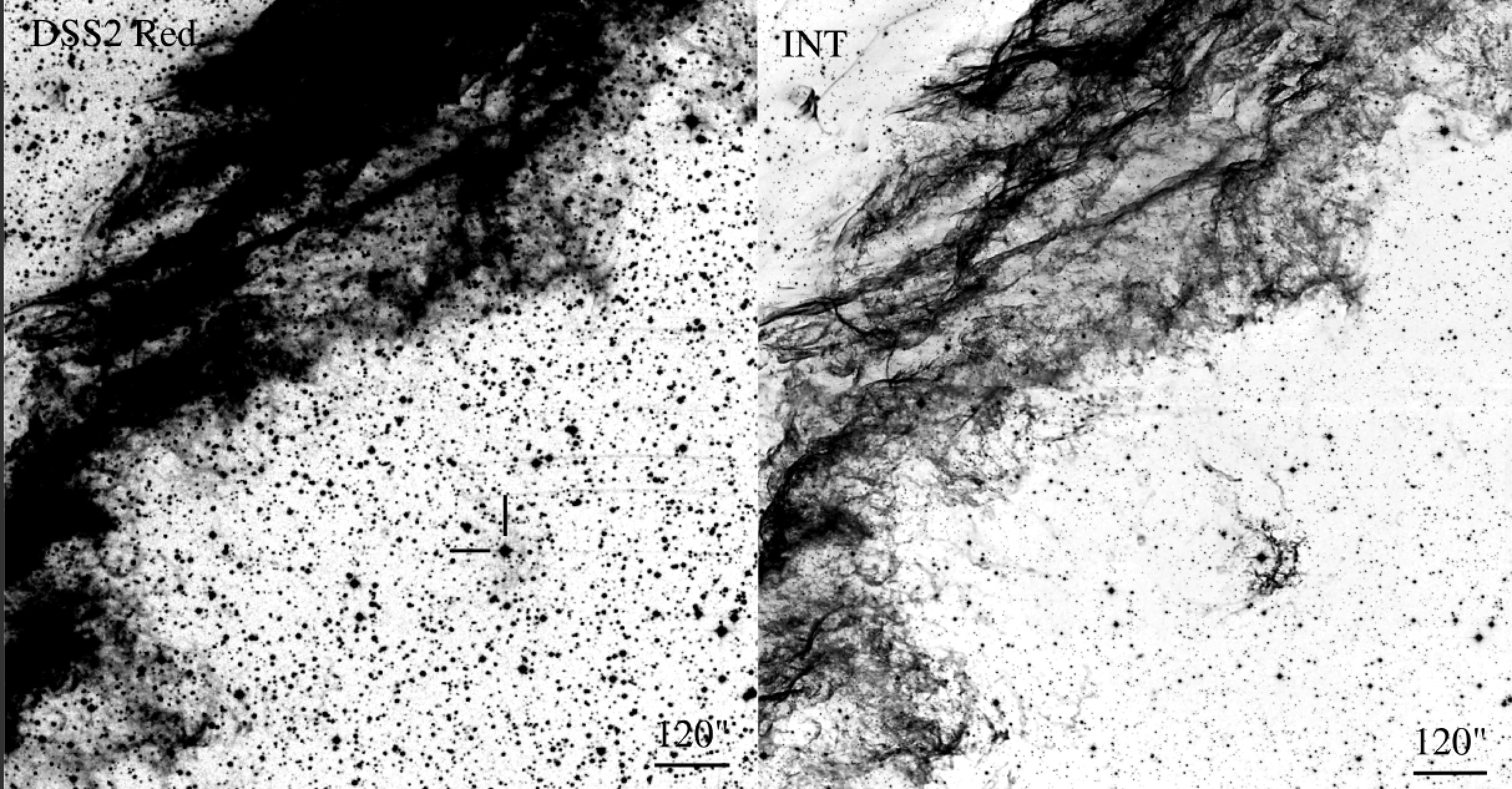}
\caption{Left panel: Enlargement of the DSS2 red image of
the south-central portion of NGC 6992
with the star J205601+313130 marked. Faint emission is seen
around the star, mainly to the west.
Right Panel: Same region imaged by Daniel Lopez using the Isaac Newton Telescope (INT)
and a combination of broad filter and narrow H$\alpha$ images which
highlights the optical nebulosity around the star. }
\label{DSS2_n_INT}
\end{figure*}

%%%%
%%%%
%  \section{Results on Stars Suspected to Lie Inside the Cygnus Loop Supernova Remnant}
\section{Results}
%%%%
%%%%

Below we present spectral and imaging data on two stars which we suspect lie
physically inside the Cygnus Loop remnant based on evidence suggesting
interactions of these stars' mass loss material with the remnant's shock front
in the form of surrounding optical nebulosities.

% Figure 4
\begin{figure*}
\begin{center}
\includegraphics[width=0.95\linewidth]{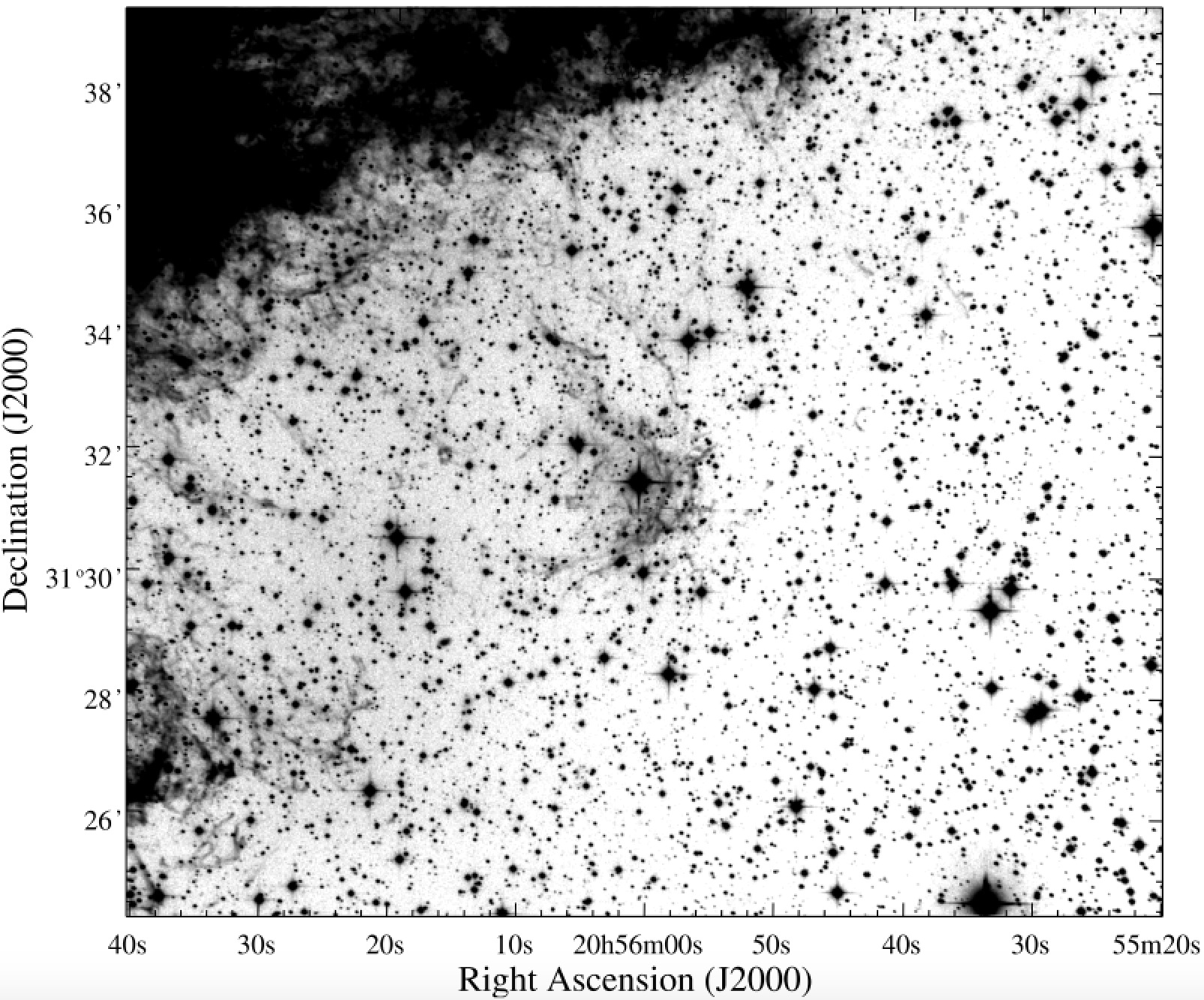}
\end{center}
\caption{H$\alpha$ image of the nebulosity around J205601+313130.
Note the diffuse emission seen to the east of J205601 which appears to extend, broaden and merge
in with the western edge of the Cygnus Loop's bright nebula NGC~6992.
North is up, East to the left.}
\label{Ha_image}
\end{figure*}

% Figure 5
\begin{figure*}
\begin{center}
\includegraphics[width=0.90\linewidth]{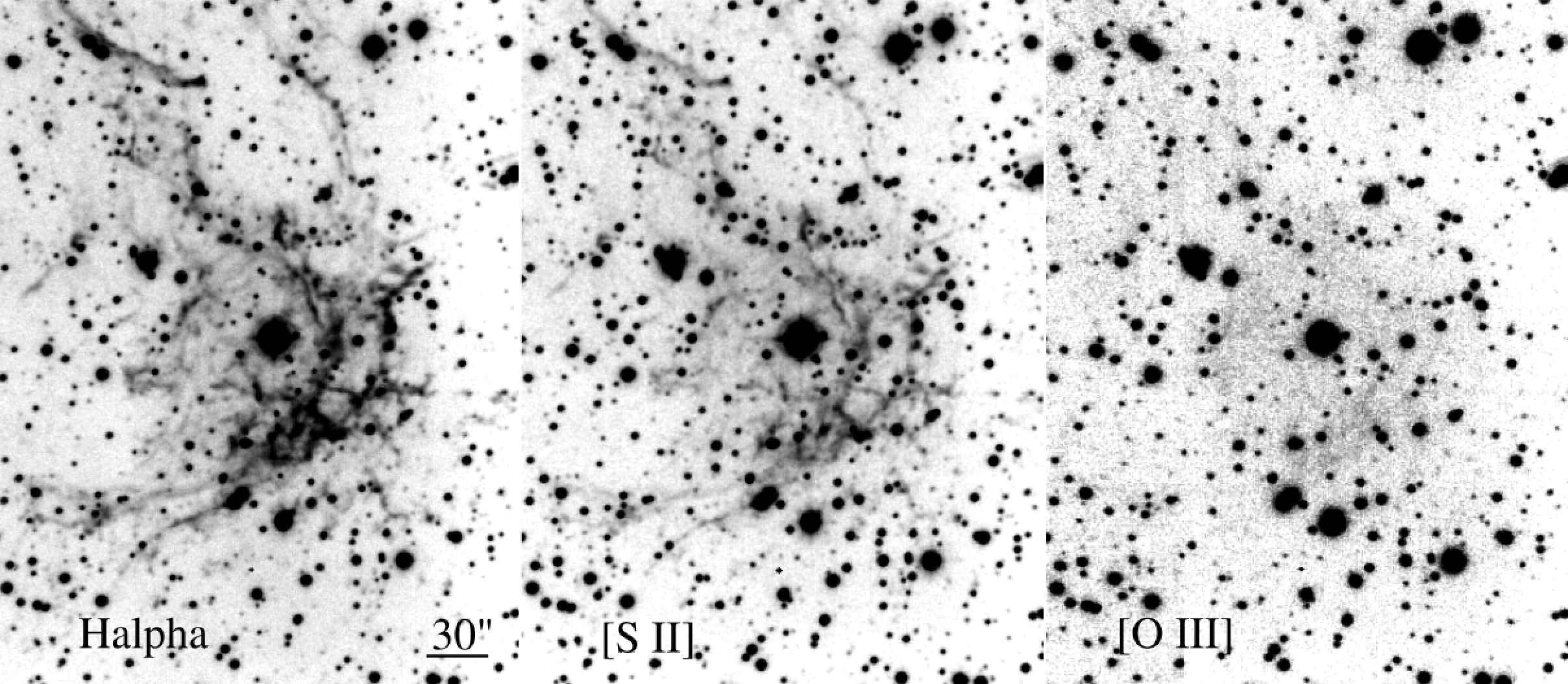}
\caption{Images of the J205601 nebulosity  taken in the light of
H$\alpha$, [\ion{S}{2}] $\lambda\lambda$ 6716,6731, and [\ion{O}{3}] $\lambda$5007.
North is up, East to the left.  }
\end{center}
\label{Mstar_neb_tiles}
\end{figure*}

% Figure 6
\begin{figure*}[t]
\begin{center}
%\plotone{Mstar_slit_positions_v1.ps}
%\includegraphics[width=0.5\linewidth]{Mstar_slit_positions_GC10.pdf}
%\includegraphics[width=0.65\linewidth]{Mstar_slit_positions_GC10.jpg}
\includegraphics[width=0.65\linewidth]{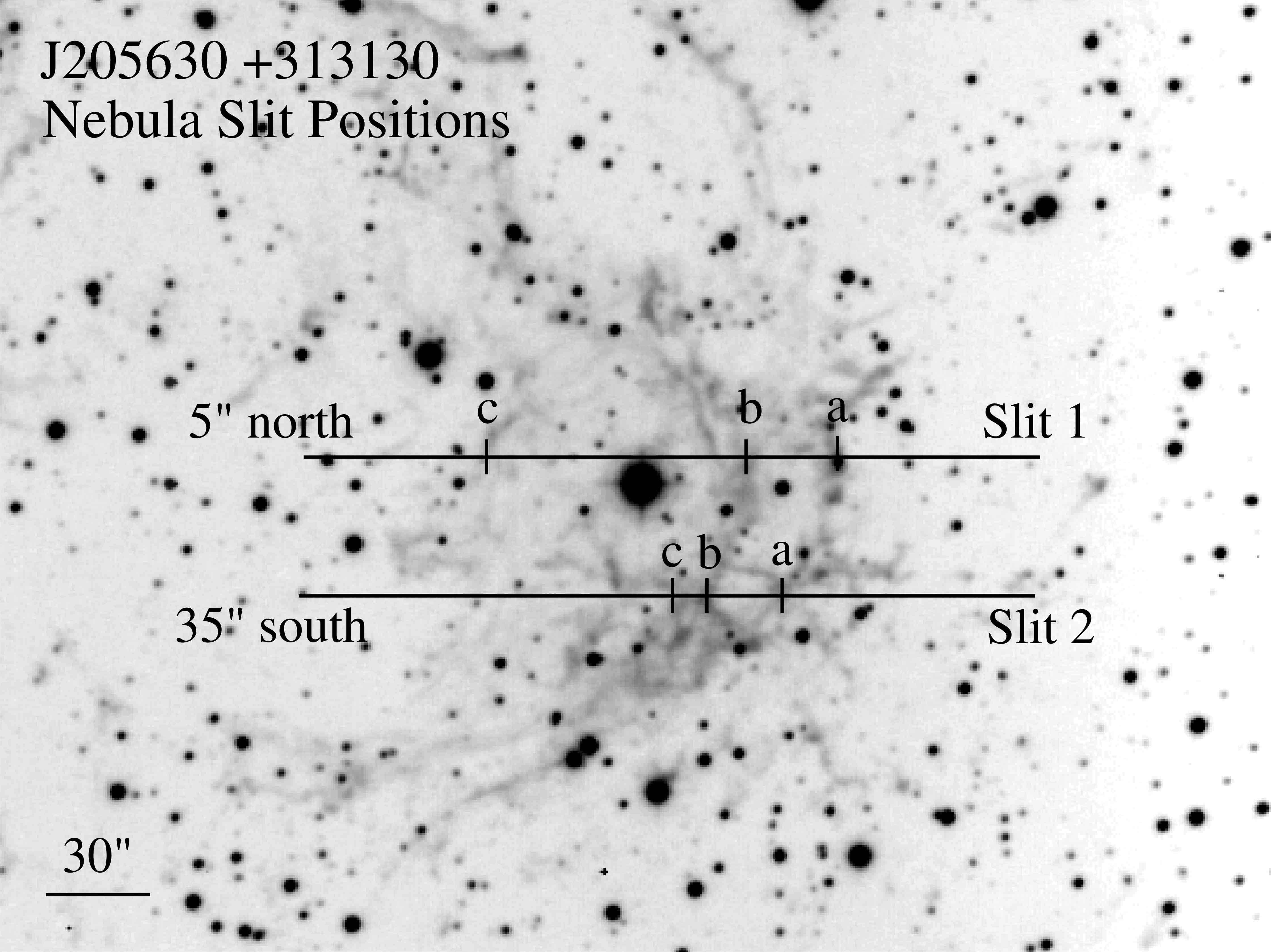}
\caption{Slit positions in the nebulosity above and below J205601+313130
where low-dispersion red spectra were obtained (see Fig.\ 7).
North is up, East to the left.  }
\end{center}
\label{slit_positions}
\end{figure*}

% Figure 7             
\begin{figure*}
\begin{center}
\includegraphics[width=0.80\linewidth]{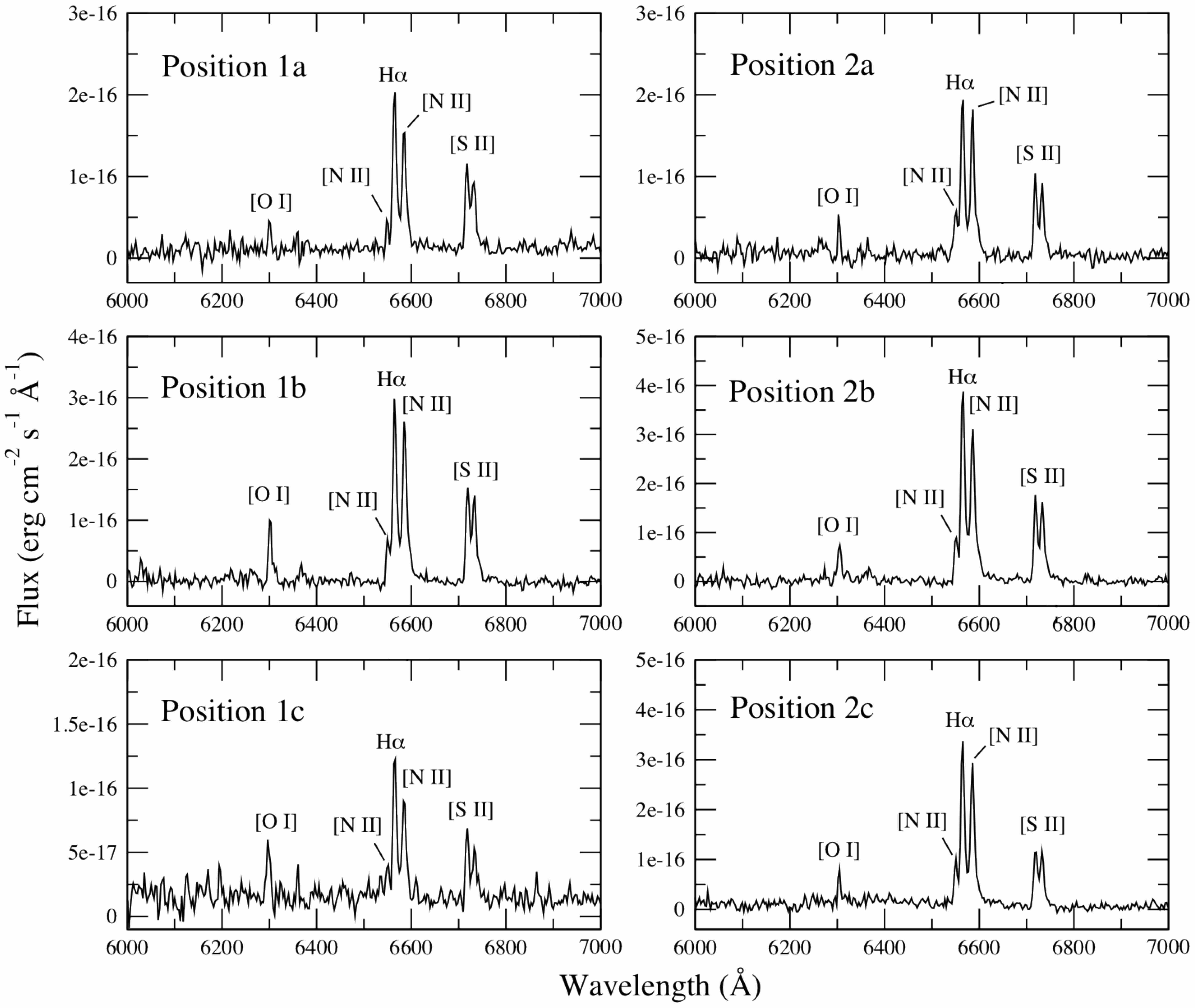}
\caption{Spectra of six positions in the nebulosity around J205601
along slits Position 1 and 2 (see Fig.\ 6).  }
\end{center}
\label{Mstar_neb_spectra}
\end{figure*}

\subsection{J205601: A M4 III Inside a Bow-Shock Nebula}

%%%%%% Table 2

%\begin{deluxetable}{lcccccc}
\begin{deluxetable*}{lrrrrrr}
\tablecolumns{7}
\tablecaption{Observed Relative Fluxes in the M Star Nebula}
\tablewidth{0.95\linewidth}
\tablehead{
\colhead{Line/Ratio} &
       \colhead{Pos 1a} & \colhead{Pos 1b} & \colhead{Pos 1c} & 
\colhead{Pos 2a} & \colhead{Pos 2b} & \colhead{Pos 2c} }
\startdata
~  [O I] 6300       & 11    & 25  & 37  & 12  & 16  & 10  \\
~ H$\alpha$ 6563    & 100   & 100  & 100  & 100  & 100  & 100  \\
~ [N II] 6583       & 80    & 98  & 84  & 92  & 95  & 96  \\
~ [S II] 6716       & 56    & 54   & 50 & 47  & 43  & 39  \\
~ [S II] 6731       & 52    & 48   & 42 & 45  & 42  & 38  \\
~ [S II]/H$\alpha$  & 1.08  & 1.02 & 0.92 & 0.89 & 0.85 & 0.77  \\
~ [S II] 6716/6731  & 1.08  & 1.12 & 1.20 & 1.04 & 1.02 & 1.03  \\
~ $\rho$ (cm$^{-3}$)& 430   & 360  & 250  & 510  & 560  & 540   \\
~ H$\alpha$ flux\tablenotemark{a}    & 1.9   & 3.2  & 1.0  & 2.1  & 4.1  & 3.4   \\
\enddata
\tablenotetext{a}{Flux units: $10^{-15}$ erg s$^{-1}$ cm$^{-2}$. }
\end{deluxetable*}

\subsubsection{Stellar Classification}

A low-dispersion spectrum of J205601, covering the wavelength region 3800 --
6000 \AA,  is shown in  Figure 2. Through a comparison with MK spectral
standards \citep{GC2009}, we find that J205601 is an M4 red giant based on the
strength of the observed TiO bands as well as the distinct features at 4626 and
4667 \AA.  In order to determine luminosity class, we looked at the strength of
the MgH band near 4770, which is prominent in main sequence M stars. Using
this, as well as direct comparisons between all M4 luminosity classes, we
conclude J205601 to be a M-giant, specifically an M4 III.

Our measured V = $11.60 \pm 0.03$ and B = $13.26 \pm 0.03$ values for J205601
are consistent with Tycho-2 catalog values \citep{Hog2000} of V = $11.57 \pm
0.12$  and B = $13.08 \pm 0.30$ (see Table 2).  After correcting the observed
$B-V$ for extinction to the Cygnus Loop of $E(B-V)$ = 0.08
\citep{Parker1967,Ray1981,Fesen1982}, the star's $B-V$ of $+1.58$ is in
line with $\simeq +1.62$ expected for an M4 III star \citep{Cox2000}. 

However, a range of $B-V$ values spanning 1.40 -- 1.65 have been observed for
M4 III stars \citep{Pickles1998} meaning there could be some extra visual
extinction possibly due to dust in the star's surrounding nebulosity. An
$E(B-V) > 0.08$  is also suggested by the fact that while its J-H and H-K
values of 0.95 and 0.30 are close to tabulated values of M4 III stars (5.10;
\citealt{Bessell1988,Ducati2001}), its V-K value of 5.74 is some 0.6 mag larger 
than expected possibly indicating an $E(B-V)$ closer to 0.20 mag. 

\subsubsection{The Surrounding Bow-Shaped Nebulosity}
  
Figure~\ref{DSS2_n_INT} shows the location of J205601 in context within the
Cygnus Loop's bright eastern nebula, NGC~6992.  The left panel is an
enlargement of the red DSS2 image centred several arcminutes west of the
bright nebula, NGC~6992. This image reveals the presence of faint emission
around J205601 roughly two arcminutes in size.  

The nebula around J205601 is more apparent in the right panel which is a
reproduction of Daniel Lopez's INT composite H$\alpha$ + broadband filter
image.  In this image, the J205601 nebulosity appears highly
filamentary, centred on and brightest to the west of J205601, and exhibiting
a strong eastward curvature suggesting a bow shock morphology.

A deeper H$\alpha$ image of the J205601 nebula is shown in
Figure~\ref{Ha_image}.  Although lying at a significant distance west of
NGC~6992, this image reveals faint emission extend eastward from the J205601
nebulosity over to the trailing edge of NGC~6992.  The faint intervening
emission is both diffuse and filamentary in places and, along with the lack of
any detectable H$\alpha$ emission west or south of J205601, suggests a physical
connection between the J205601 nebula with NGC~6992, thus implying that they
lie at the same approximate distance. 

The J205601 nebula exhibits strong [\ion{S}{2}] line emission relative to
H$\alpha$, like that observed in supernova remnants and the majority of
Cygnus Loop filaments \citep{Fesen1982}.  This can be seen in Figure 5 which shows
the J205601 nebulosity in the emission lines of H$\alpha$, [\ion{S}{2}] 6716,
6716 \AA, and [\ion{O}{3}] 5007 \AA. 

In the H$\alpha$ and [\ion{S}{2}] images, the nebula appears to consist of two
broken concentric rings of filaments centred on J205601. The
filaments open to the north and south of the star and exhibit a
morphology not unlike that of a wind-swept nebula.  In contrast, little or no filament
emission is seen in the [\ion{O}{3}] image, where only faint, diffuse emission
centred on the star is visible which may be, in part, reflections of J205601 in the narrow
passband filter. Comparison of DSS1, DSS2, and our more recent images reveals
significant eastward proper motion of the nebula's filaments somewhat smaller in magnitude
to that seen in the neighboring NGC~6992 filaments.

Low-dispersion spectra of six locations in the J205601 nebula were taken using
two E-W slits, as shown in Figure~6. The resulting spectra are shown in Figure
7 with observed relative fluxes are listed in Table 3.

These spectra clearly indicate that the nebula consists of shocked material.
The relative strength of [\ion{S}{2}]/H$\alpha$ at all six locations is greater
than the 0.4 criteria for identifying optical shocked material
\citep{MC72,Raymond1979,Shull1979,Dodorico80,Dopita84}.  Moreover, the absence of
appreciable [\ion{O}{3}] 4959, 5007 emission in the nebula (see
Fig.~\ref{Mstar_neb_tiles}) indicates a relatively low shock velocity.  Shock
models of interstellar shocks show that [\ion{O}{3}] line emissions only become
strong at velocities above 90 km s$^{-1}$
\citep{Shull1979,Raymond1979,Sutherland2003}. A relatively low shock velocity
in the 205601 nebula is in contrast with the majority of Cygnus Loop filaments,
possibly the results of greater electron densities. 

In fact, estimated electron densities in the J205601 nebula, based on the density
sensitive [\ion{S}{2}] 6716/6731 ratio, appears significantly higher than that
seen generally in the Cygnus Loop's filaments.  In a study of
several of the remnant's brighter filaments, \cite{Fesen1982} found
[\ion{S}{2}] 6716/6731 values from 1.30 to 1.45, with the majority at or close
to the low density of 1.43 \citep{Osterbrock} indicating electron densities
$\leq 100$ cm$^{-3}$ assuming an electron temperature of $10^{4}$ K.  Similar
[\ion{S}{2}] 6716/6731 lines above 1.30 were also found in an isolated ISM
cloud in the SW region of the Cygnus Loop \citep{Patnaude2002}.
 
In contrast, the observed [\ion{S}{2}] 6716/6731 line ratio for the six regions
in the J205601 nebula show ratios from 1.02
to 1.20 indicating electron densities of roughly 250 - 550 cm$^{-3}$ (Table
3).  Although [\ion{S}{2}] 6716/6731 line ratios around 1.2 were reported by
\citet{Fesen1996} for two Cygnus Loop filaments, it appears that electron
densities in the J205601 nebula are much higher than the majority of filaments in the
Cygnus Loop remnant.

% Figure 8
\begin{figure*}
\begin{center}
\includegraphics[width=0.90\linewidth]{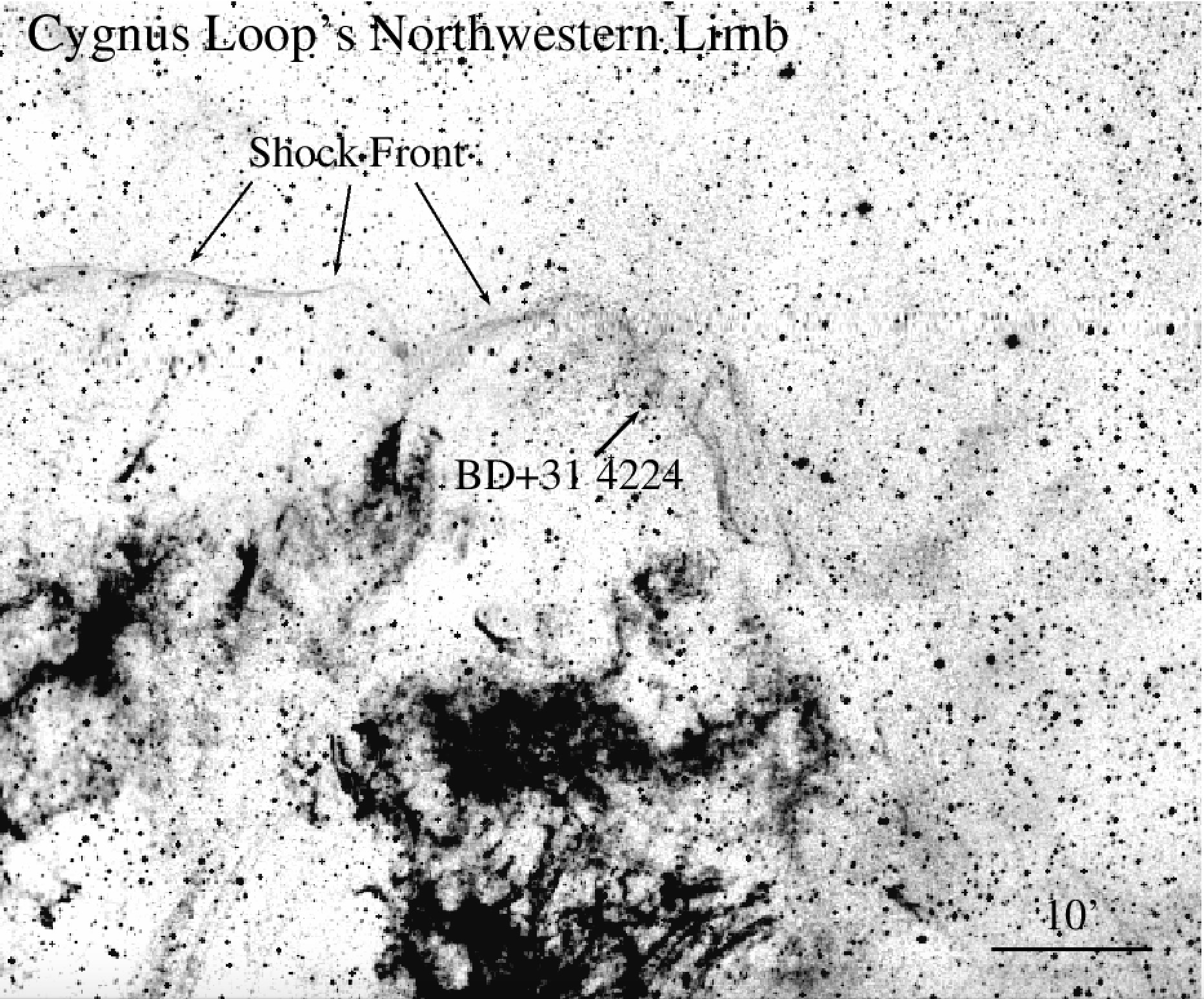}
\caption{Wide field Schmidt H$\alpha$ image of the northwestern limb of the Cygnus Loop
showing disturbances of the remnant's forward shock front, one in the region
near BD+31 4224.
North is up, East to the left.  }
\end{center}
\label{Bstar_Schmidt}
\end{figure*}

% Figure 9
\begin{figure*}
\begin{center}
%\plotone{fig_cygnus_flat.eps}
%\includegraphics[width=1.0\linewidth]{fig_cygnus_bstarspecGC10.pdf}
%%%
%\includegraphics[width=0.95\linewidth]{cygnus_bcurve_edit_vG10.jpg}
%\includegraphics[width=0.95\linewidth]{cygnus_bflat_edit_v1_G10.jpg}
%\includegraphics[width=0.95\linewidth]{f9a.jpg}
%\includegraphics[width=0.95\linewidth]{f9b.jpg}
\includegraphics[width=0.95\linewidth]{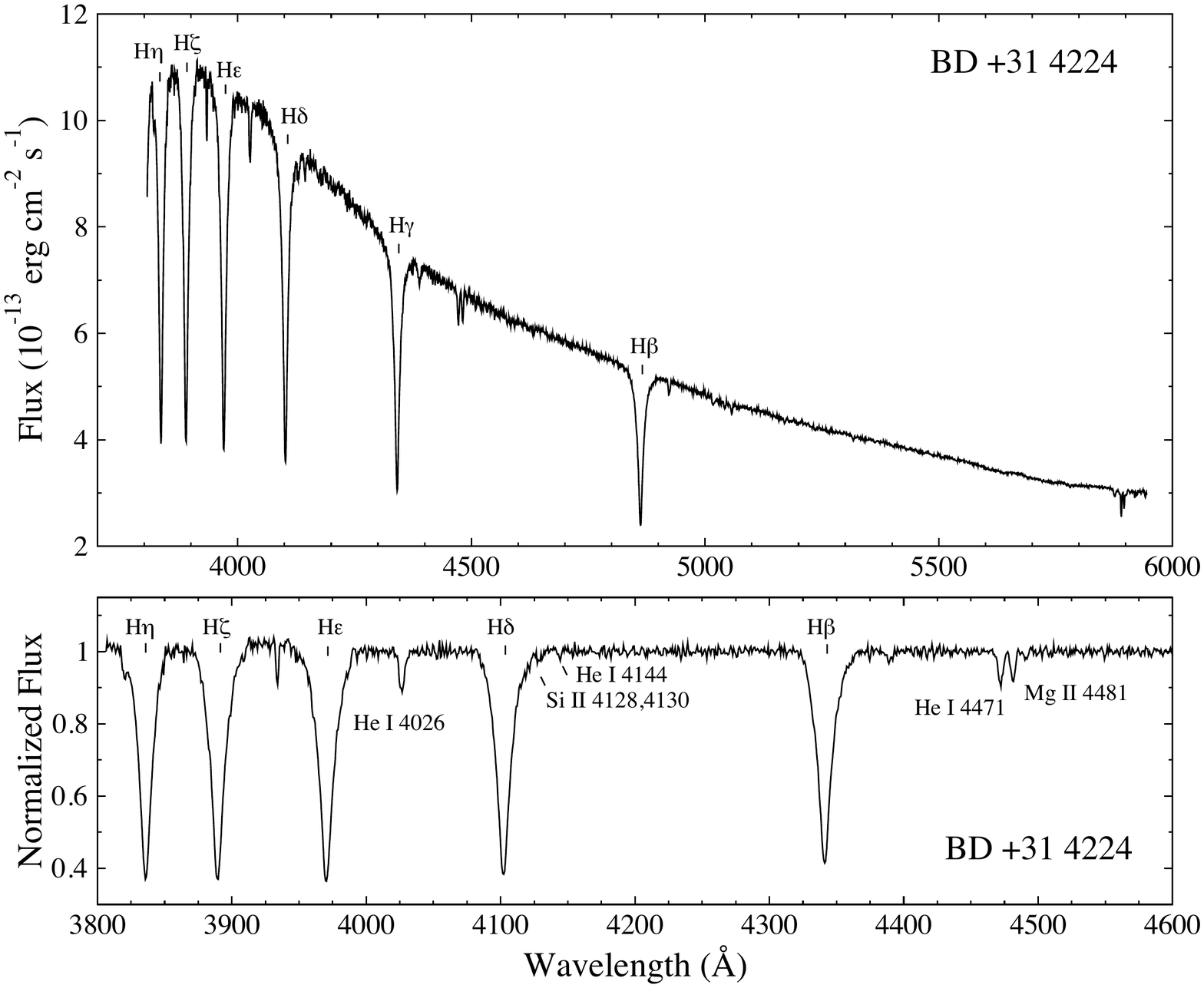}
\caption{Observed (top) and normalized (bottom) spectrum of BD+31 4224.}
\end{center}
\label{Bstar_normal_spectrum}
\end{figure*}

\subsection{BD+31 4224: A B7 Star in a Crescent Nebula}

A wide-angle H$\alpha$ image of the northwestern limb of the Cygnus Loop is
shown in Figure 8. This image is shown to highlight disturbances in the
remnant's forward shock front along the remnant's northernmost boundary.  
One shock front disturbance can be  
seen associated with a small interstellar cloud, visible just left of image centre,
and another shock front disturbance farther to the west near the star BD+31~4224.

It was not the presence of either disturbance in the remnant's northern shock front that
first drew our attention to BD+31~4224. Rather, the star's location relative to a small
crescent-shaped nebula (see below) that is only weakly visible on the red
DSS2 image that led to an investigation into the nebula's nature and subsequently to this
star.   

% Figure 10 
\begin{figure*}
\begin{center}
%\ncludegraphics[width=0.95\linewidth]{Bstar_FOV.jpg}
%\includegraphics[width=0.95\linewidth]{Bstar_Halpha_n_sub_v2_G10.jpg}
%%
\includegraphics[width=0.95\linewidth]{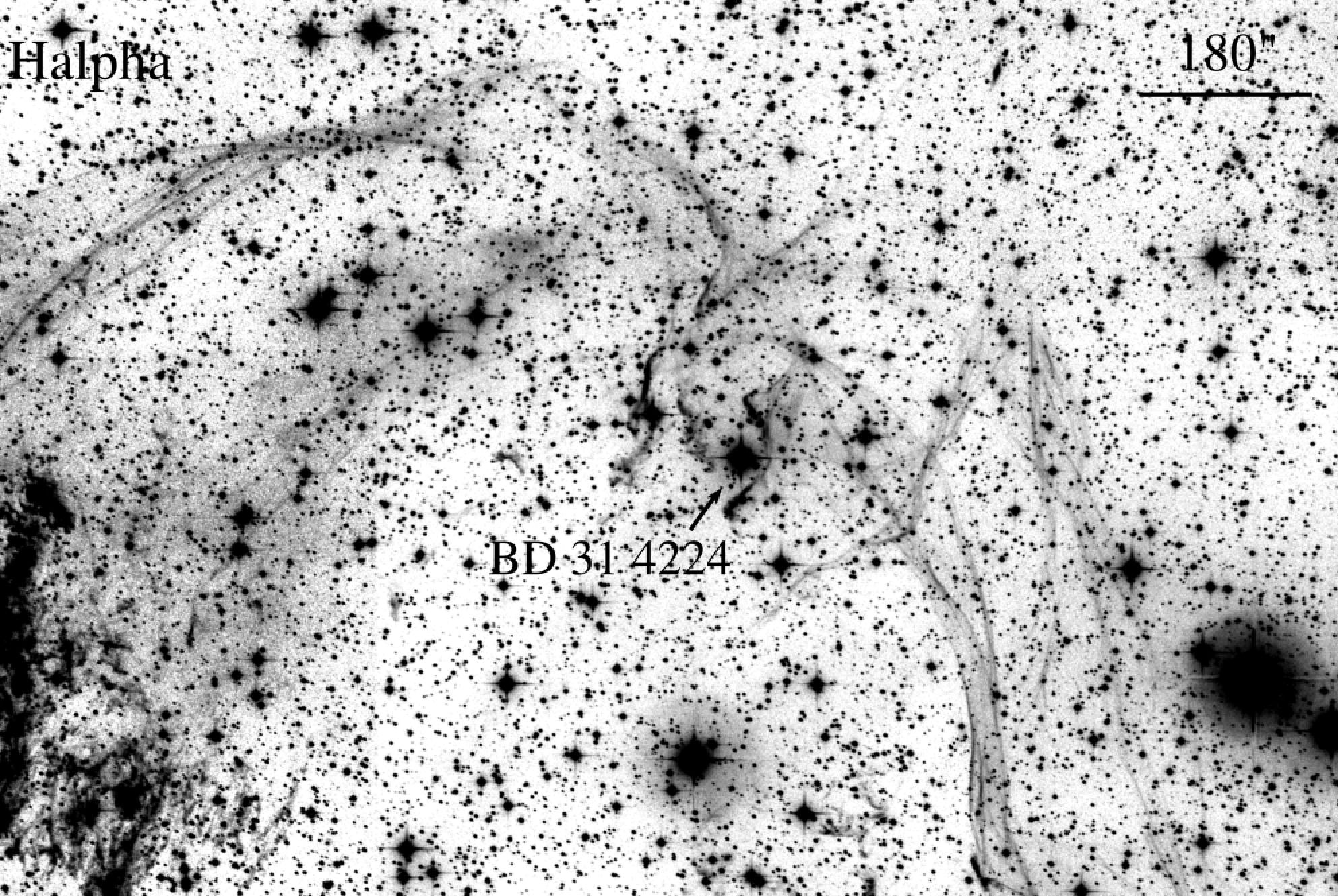}
\includegraphics[width=0.95\linewidth]{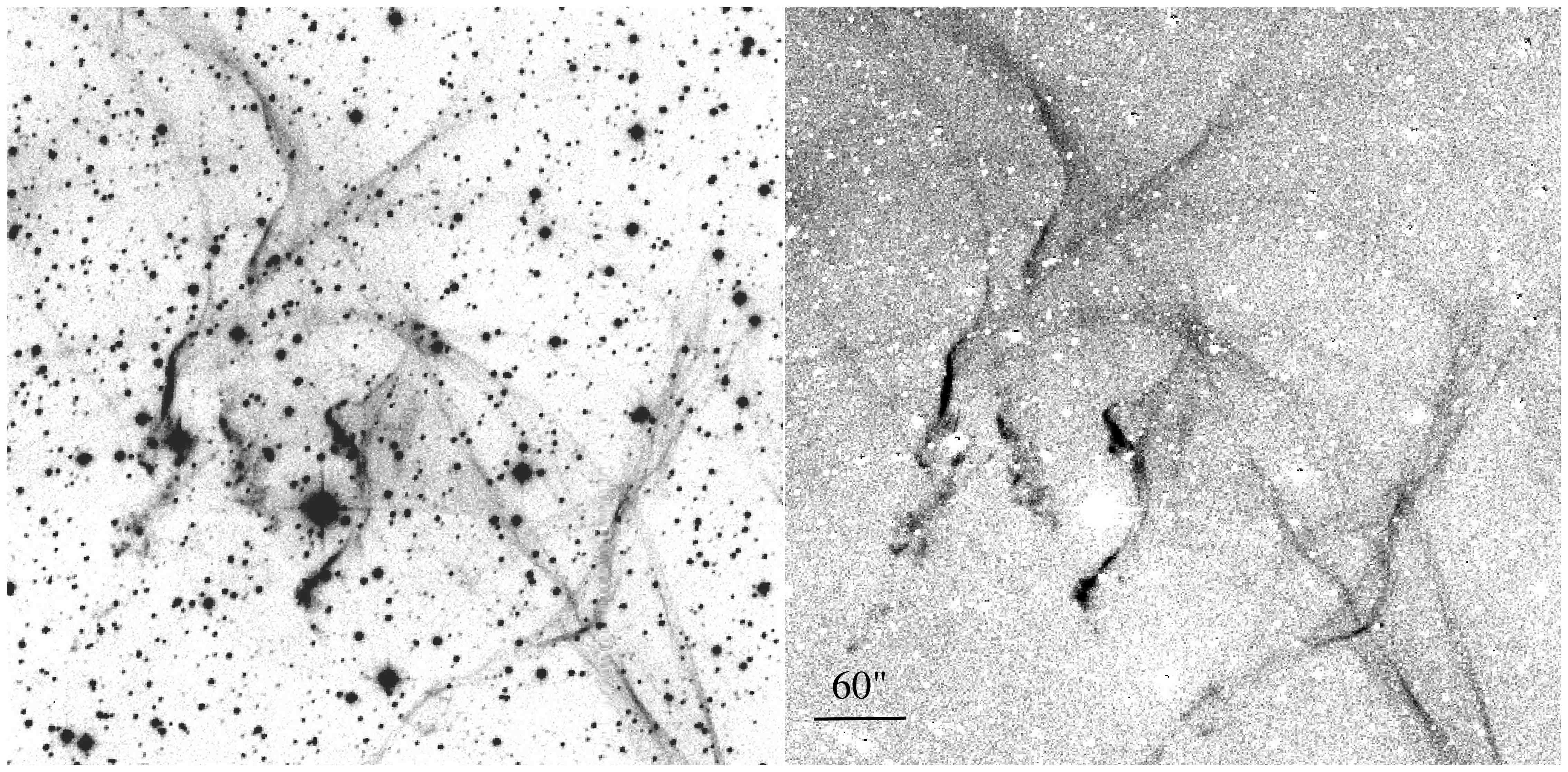}
\caption{Top Panel: Deep H$\alpha$ image of the emission structure
along the Cygnus Loop's northeastern limb and centred on
the B star BD+31 4224. North is up, East to the left.
Bottom Panels: Enlargement of the H$\alpha$ image with stars (left) and with
stars partially removed (right) to better reveal the filaments and
nebulosity around BD+31~4224.    }
\end{center}
\label{Bstar_wideview}
\end{figure*}

% Figure 11
\begin{figure*}
\begin{center}
%\plotone{Bstar_tile_v3_600.ps}
%\includegraphics{Bstar_tile_v3_600_GC10.pdf}
%%
%\includegraphics{Bstar_tile_v3_600_GC10.jpg}
%%
\includegraphics[width=0.90\linewidth]{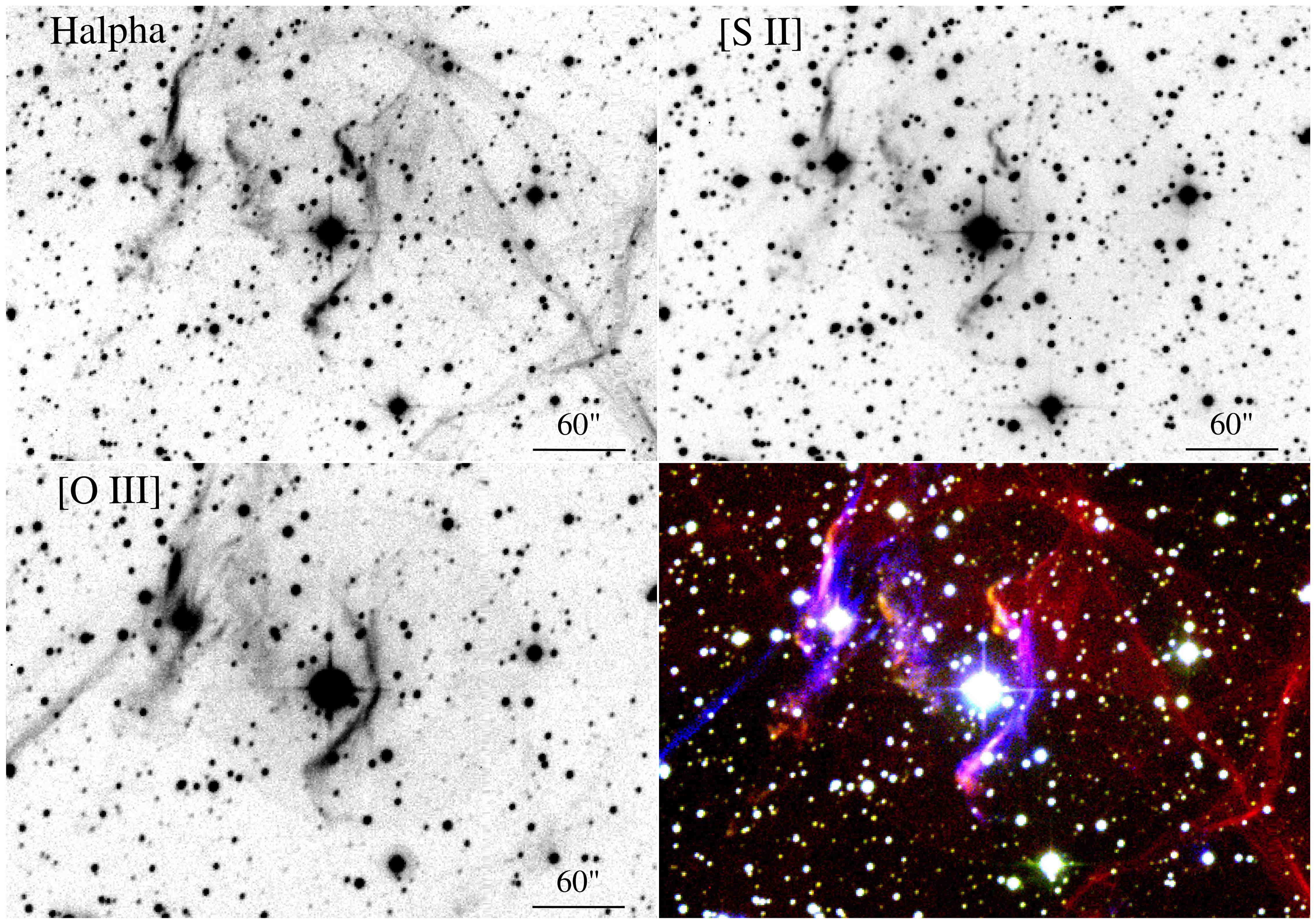}
\caption{H$\alpha$, [\ion{S}{2}] 6716,6731 and [\ion{O}{3}] 5007 images
of the emission structure around the B star BD+31 4224 along with a color composite
(H$\alpha$ red,  [\ion{S}{2}] green, \& [\ion{O}{3}] blue).
Note the similarity of the H$\alpha$ and [\ion{S}{2}] images
and the contrast with that of the [\ion{O}{3}] image
especially the centre of the emission arc nearest the B star.
North is up, East to the left. }
\end{center}
\label{Bstar_tile}
\end{figure*}

% Figure 12
\begin{figure*}
\begin{center}
%\plotone{fig_cygnus_bstarslit.eps}
%\includegraphics{Bstar_slit_positions_coordGC10.pdf}
%%
%\includegraphics[width=0.90\linewidth]{Bstar_slit_positions.jpg}
%%
\includegraphics[width=0.7\linewidth]{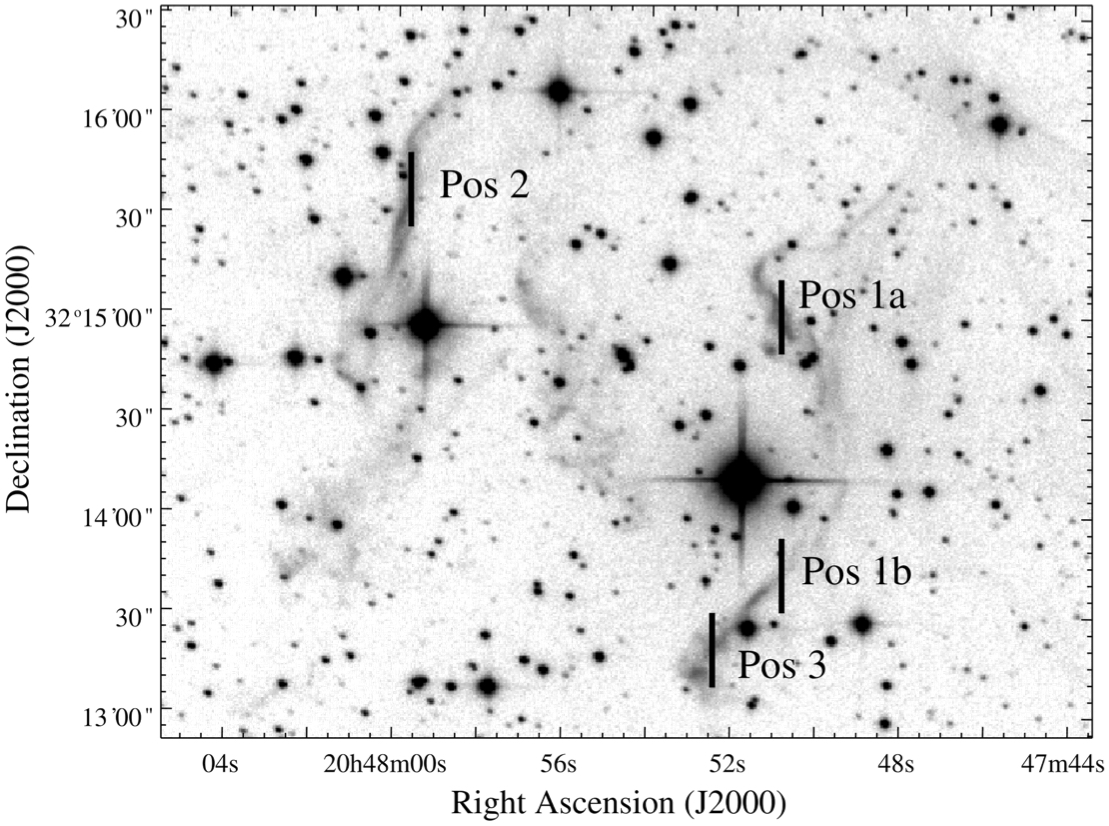}
\caption{Slit positions in the nebulosity around the B star BD+31 4224
where low-dispersion red spectra were obtained (see Fig.\ 12).
North is up, East to the left.  }
\end{center}
\label{B_star_slit_positions}
\end{figure*}

% Figure 13
\begin{figure*}
\begin{center}
%\includegraphics[angle=-90,width=1.00\linewidth]{fig_cygnus_filspec.eps}
%\includegraphics{B_nebula_spectra.pdf}
%%%%
%\includegraphics[width=1.0\linewidth]{cygnus_spectra_edit_v2_G10.jpg}
%%%%
\includegraphics[width=1.0\linewidth]{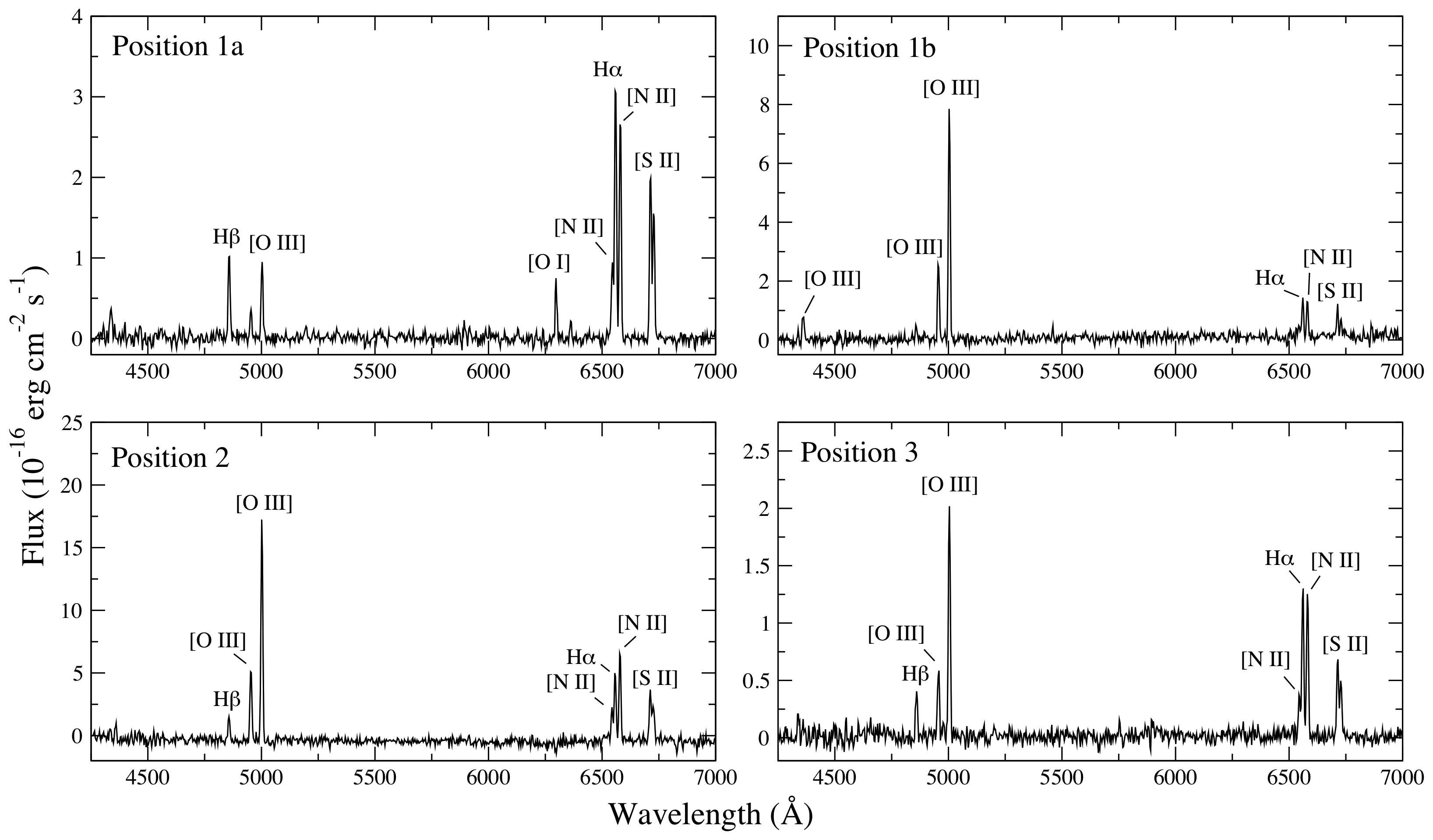}
\caption{Spectra of filaments near BD+31 4224. See Figure 12 for slit positions. }
\end{center}
\label{Bstar_spectra}
\end{figure*}

\subsubsection{Stellar Classification}

A low-dispersion spectrum of BD+31~4224 covering the wavelength region 3800 --
6000 \AA \  indicates it is a late B type star for which we have determined a spectral
type of B7.  As done an aid in spectral classification, we flattened its 
spectrum and the normalized flux plot of its spectrum is shown in
Figure~9. This figure shows the important lines used in determining the
spectral type, namely Si II~4128, 4130 \AA, He I~4144,4471 \AA, and Mg~II 4481
\AA. He~I 4471/Mg~II 4481 serves as a ratio to indicate late spectral types.

The classification of B7 is mainly on the basis of the comparison of
\ion{He}{1} 4471 \AA \ with \ion{Mg}{2} 4481 \AA, and the weakness of the
\ion{He}{1} 4144 which fades in late B stars \citep{GC2009}.  Although the
Si~II 4128, 4130 \AA\ features are somewhat obscured by the wide H$\delta$ at
4102 \AA, it appears that the Si~4128, 4130 \AA\ lines are stronger than He I
4144 \AA, firmly identifying it as B7. 

Because of the factors listed above and the fact that 
it exhibits a spectrum virtually identical to the primary B7 V standard
star standard HR 1029 (= HD 21071) \citep{Keenan1985,Garcia1989,GG1994,GC2009},
a spectral classification of B7 V appears secure.  In addition, the star's
observed extinction corrected $B-V$ value of $-0.12$ is consistent with
standard B7 V $B-V$ values of $-0.12$ to $-0.13$
\citep{Pickles1998,Cox2000,mamajek2017}.

Although a dwarf luminosity classification (V) is indicated based on the
broadness of its hydrogen lines, we could not completely rule out a possible
subgiant IV classification based upon the weakness of \ion{O}{2} 4070 and 4416
\AA \ lines due to our spectrum's relative low S/N.

Our photometric observations of this star agree with values from the Tycho-2
catalog \citep{Hog2000}.  Specifically, we measured V = $9.59 \pm 0.02$ and B =
$9.55 \pm 0.03$ which are close to the listed Tycho-2 values of V = 9.58 and B
= 9.53 (see Table 2).   

\subsubsection{Nebulosities Near BD+31~4224}

The faint nebulosities near and around BD+31~4224 are better seen in Figure 10.
A relatively deep H$\alpha$ image is presented in the top panel, with
enlargements of an area immediately around BD+31~4224 shown in the bottom
panels. As seen in these images, H$\alpha$ emission filaments associated with
the Cygnus Loop's forward shock exhibit a strong distortion at its
northernmost boundary near BD+31~4224. Specifically, a ``blow-out'' of the remnant's shock
front can be seen along with several highly curved H$\alpha$ filaments with
BD+31~4224 appearing symmetrically placed relative to these emission features.  

On smaller scales, BD+31~4224 lies centred within a bow-shaped emission arc. Figure 11
shows images of this emission arc taken in H$\alpha$, [\ion{S}{2}] 6716, 6731 \AA \
and [\ion{O}{3}] 5007 \AA. Noticeably stronger H$\alpha$ and [\ion{S}{2}] emission
is seen at the northern and southern tips of the bow-shaped nebula, whereas
[\ion{O}{3}] is strongest in its middle and closest to BD+31~4224. 

Emission can be seen to extend about 2 arcminutes to the northeast of
BD+31~4224, especially prominent in the [\ion{O}{3}] image, and near a noticeable 
but fainter star compared to BD+31~4224 (m$_{V}$ = 12.0; TYC 2691-1550-1). A
spectrum shows it to be a K dwarf and it has a Tycho-GAIA determined parallax
measurement of $7.70 \pm 0.29$ \citep{Gaia2016}. Its optical spectrum, observed
magnitude, and measured parallax indicates it lies at $\sim$ 130 pc and is thus
unrelated to the Cygnus Loop and the observed emission filaments.   

Optical spectra were obtained at three positions in the bright nebula arc west
of BD+31~4224, plus one position in a filament of similar brightness
located some $150''$ northeast of BD+31~4224 (see Fig.\ 12). The resulting
spectra of these four regions are shown in Figure 13 with observed
relative line strengths listed in Table 3.

The observed strength of the [\ion{S}{2}] 6716, 6731 \AA \ lines relative to that of
H$\alpha$ clearly indicate that all four regions represent shock emission.  The
spectrum of the filament off to the NE of BD+31~4224 is not markedly different
from that seen in the bow-shaped arc nebula centred on BD+31~4224.  However,
unlike that seen in the J205601 nebula where the density sensitive [\ion{S}{2}]
lines indicate electron densities above 250 cm$^{-3}$, the shocked nebulosities
around BD+31~4224 are much lower $\lesssim 100$ cm$^{-3}$ 
near the low density limit of 1.43 \citep{Osterbrock}.

The strength of the [\ion{N}{2}] 6548, 6583 \AA \ lines relative to that of
H$\alpha$ are relatively strong but not unusual for SNRs and in particular
Cygnus Loop filaments \citep{Fesen1982,Fesen1996}. The same holds true for the
[\ion{O}{3}] lines. However, there is a significant difference in the
[\ion{O}{3}] strength at either end of the bow-shaped nebula (Pos 1a and Pos 3)
versus Pos 1b, closer to the B star BD+31~4224. This is consistent with the
narrow passband images shown in Figure 11.

The location of BD+31~4224, a B7 star which is expected to have relatively
high-velocity stellar winds, at the centre of a small bow-shaped nebula,
brightest in [\ion{O}{3}] closest to the star, and symmetrically placed in
relation to a series of highly curved filaments and situated near a shock front
blow-out along the remnant's northern shock boundary together comprise strong
evidence for a physical connection of the star's stellar winds and the Cygnus
Loop's shock wave.  In this scenario, the  B7 star's winds interacted with the
remnant's shock front creating both the observed bow-shock nebula centred on the
star and the distorted filaments and shock front break-out in this
northwestern limb region. 

%******************************************************************************
%******************************************************************************

\section{Discussion}

Underlying every quantitative discussion of the Cygnus Loop is uncertainty of 
its distance. As shown in Table 1, prior distance estimates have ranged from
less than 0.4 to nearly 2.0 kpc. Because of the remnant's prominent place in the
study of evolved Galactic SNRs at all wavelengths, it is important to
determine its distance to a greater degree of certainty. Below, we first review
some of the previous distance measurements and then describe how the two stars
discussed above may help resolve this issue.

%%%%%% Table 3
\begin{deluxetable}{lrrrr}
%\scriptsize
\tablecolumns{5}
\tablecaption{Relative Fluxes in Nebula Near BD+31~4224}
%\tablewidth{0.8\linewidth}
\tablehead{
\colhead{Line/Ratio} & \colhead{Pos 1a} & \colhead{Pos 1b} & \colhead{Pos 2} & \colhead{Pos 3} }
\startdata
~  [O III] 4363   &\nodata  & 25   & --   &  \nodata   \\
~ H$\beta$ 4861      & 100  & 100  & 100  &  100  \\
~  [O III] 5007      &  90  &2020  & 850  &  475      \\
~ H$\alpha$ 6563     & 315  & 300  & 285  &  315  \\
~ [N II] 6583        & 265  & 280  & 355  &  305  \\
~ [S II] 6716        & 210  & 225  & 260  &  175     \\
~ [S II] 6731        & 175  &(142) & 175  &  137    \\
~ [S II]/H$\alpha$   & 1.23 & 1.23 & 1.52 &  1.00     \\
~ [S II] 6716/6731   & 1.20 &(1.6) & 1.48 &  1.28     \\
~ $\rho$ (cm$^{-3}$) & 250  &$<$100&$<$100&  150      \\
~ H$\alpha$ flux\tablenotemark{a} & 3.5  & 1.2 & 5.8  & 1.4   \\
\enddata
\tablenotetext{a}{Flux units: $10^{-15}$ erg s$^{-1}$ cm$^{-2}$. }
\end{deluxetable}

\subsection{Previous Distance Estimates to the Cygnus Loop}

Until the mid-1970's, the most widely adopted value for the Cygnus Loop's
distance was from a kinematic investigation by \citet{Minkowski1958}.  This
approach has the inherent uncertainty of connecting proper motions seen in one
set of filaments with the radial velocities or inferred shock velocity found
for some other filaments. 

Consequently, Minkowski's Cygnus Loop distance
estimate began to be questioned. \citet{McKee1975} suggested 770 pc might be an
underestimate while \citet{Kirshner1976} thought it was much too large.
Subsequently, a number of distances estimates were proposed usually relying on
combining proper motion with shock velocity measurements or estimates (see
Table 1).

\citet{Blair1999} compared the locations of non-radiative filaments along the
remnants northeastern limb on a digitized version of the 1953 DSS1 red plate of
the Cygnus Loop and on a 1997 Wide Field Planetary Camera2 (WFPC2) {\sl Hubble
Space Telescope} H$\alpha$ image to deduce a value of $440^{+130}_{-100}$ pc.
Later measurements of these same filaments based solely on {\sl HST} images
taken 4 years apart resulted in a revised value of $540^{+130}_{-80}$ pc
\citep{Blair2005}.  

An upper limit to the distance to the Cygnus Loop was later proposed based on 
optical and far UV observations of sdOB star lying in the direction to Cygnus
Loop's eastern NGC~6992 nebulosity which showed broad O~VI $\lambda$ 1032 line
absorption indicating it lies behind the remnant \citep{Blair2009}. Model fits
to this star's optical and UV spectra yielded a $T_{eff} = 35,500 \pm 1000$ K
and a distance of $576 \pm 61$ pc, a value consistent with the earlier remnant
estimate by \citet{Blair2005}.

More recently, \citet{Medina2014} obtained high-resolution Echelle
spectra of faint Balmer-dominated H$\alpha$ filaments along the remnant's
northeastern limb to estimate shock velocities of around 400 km s$^{-1}$ from
the broad H$\alpha$ emission component. They then combined this value with
proper motions measured by \citet{Salvesen2009} over a 39 year time span to
deduce a distance to the Cygnus Loop of $\sim$890 pc (786 -- 1176 pc). A
follow-up analysis of thermal equilibrium in a collisionless shock affecting
the broad H$\alpha$ component led to a reduction of the derived shock velocity
down from 400 to $\sim$360 km s$^{-1}$, which thereby decreased the Cygnus
Loop's estimated distance from 890 to 800 pc \citep{Raymond2015}.

While higher shock velocity estimates associated with the remnant's faint,
outer Balmer-dominated H$\alpha$ filaments may explain the relatively low
earlier Cygnus Loop distance estimates around 500 and 600 pc, the detection of
O~VI $\lambda$ 1032 line absorption in an sdOB star found by \citet{Blair2009}
along the remnant's eastern limb at an estimated distance of just $\sim575$ pc
is puzzling. 

Their spectral analysis of both UV and optical spectra of this
star by \citet{Blair2009} was thorough and robust, implying an $M_{V}$ =
$+5.23$ for this $m_{V}$ = 14.12 sdOB star.  However, some hot sdOB stars with
$T_{eff} \approx$ 35,000 K similar to this sdOB star exhibit $M_{V} = +2.5$ to
+4.5 \citep{deBoer1997}, which is much brighter that than estimated by
\citet{Blair2009}. This raises at least the possibility of a larger distance to
this star than the analysis of its spectral properties might indicate.  

\subsection{Problems with Distances Less Than 650 pc} 

\citet{Salvesen2009} investigated the
ratio of cosmic ray pressure to gas pressure in several regions in the remnant.
They constrained the shock speeds of 18 non-radiative filaments through proper
motion measurements seen along the remnant's northeastern limb from Digital Sky
Survey I and II images and adopting a distance of 637 pc, the maximum allowed
by the sdOB star observation of \citet{Blair2009}. \citet{Salvesen2009} then
deduced post-shock electron temperatures from spectral fits to {\sl ROSAT} PSPC
observations along the perimeter of the remnant and found that in most cases
the this ratio was either low or formally negative even when adopting the
maximum distance estimated of $640$ pc by \citet{Blair2009}.  

\citet{Salvesen2009} concluded the cause for the many implausible negative
ratios calculated was the significant uncertainty in the postshock temperature
measurements. However, they also wondered if the Cygnus Loop's distance might
be larger than $650$ pc, but thought it was unlikely to be as large as the
$\sim$ 1 kpc that would be needed to make all the upper limits to
$P_{CR}$/$P_{G}$ positive.

Interestingly, it has been long known that models of the Cygnus Loop remnant
assuming distances less than 1 kpc result in an energetically weak SN
explosion. Based on the framework of the Sedov model, the estimated remnant's
energy using Minkowski's 770 pc distance estimate is just E$_0$ = $3 - 7 \times
10^{50}$ (d$_{\rm pc}$/770)$^{5/2}$ erg, considerably less than the canonical SN
explosion energy of $ 1 - 2 \times 10^{51}$ erg
\citep{Rapp1974,Falle1982,Ballet1984,MT1999} with smaller distances of around
0.5 -- 0.6 kpc implying an even weaker SN event.

The problem of an energetically weak SN for the Cygnus Loop was highlighted in
a recent modeling study by \citet{PM2011}. He examined the global
parameters of the remnant assuming a low-density cavity model and using a
time-dependent spherically-symmetric hydrodynamical code and found that a
distance of 540 pc resulted in an estimated supernova explosion energy of just
$6 - 8 \times 10^{49}$ erg. A similar analysis undertaken by \citet{Fang2017}
who estimated a somewhat higher energy of $ 2 \times 10^{50}$ erg. But both estimates
fall well short of a value near $10^{51}$ erg. 

\citet{PM2011} concluded that if the Cygnus Loop were at a distance of $\approx
0.6$ kpc it must have been an unusually weak core-collapse event, perhaps ``the
weakest known core-collapse SN in the Galaxy''.  He further noted that much greater
distances, $\sim 1.25$ kpc, would be needed to recover a ``standard'' SN energy
$E_{0} \sim 1 \times 10^{51}$ erg but viewed this unlikely as it was well 
outside quoted uncertainties of the \citet{Blair2009} estimate.

\subsection{A Distance Based on Stellar Distances}  

The results of our imaging and spectroscopic investigations of nebulosities
seen around the M4 III star J205601 and the B7 V-IV star BD+31~4224 presented
in $\S$3 suggest they are circumstellar features resulting from interaction of
stellar mass loss material with the remnant's expanding shock front. If true,
this means that both stars lie inside the remnant and thus can be used to
estimate the remnant's distance using spectroscopic parallax derived distances.   

We can estimate the distances to these two stars using the familiar
distance modulus equation

\begin{equation}
log~d_{pc} = (m_V - M_V + 5- A_V)/5
\end{equation}

where $A_V = R(V)\times E(B-V)$ is the extinction. 
Using the color-excess estimate for the Cygnus Loop, 
$E(B-V) = 0.08$ \citep{Parker1967,Ray1981,Fesen1982} and 
$R(V) = 3.1$, then $A_V = 0.25$.

Absolute visual magnitude values red giants can span a wide range.
\citet{GC2009} list average $M_{V}$ values for M4 III stars between $-1.1 $ and
$-2.2$ corresponding to luminosity classes IIIa and IIIb consistent with
globular cluster measurements, while general reference sources cite values for
an M4 III star, like J205601, to be $-0.4$ to $-0.6$ \citep{Lang1992,Cox2000}.

However, {\sl Hipparcos} measurements of field stars indicate an even broader range of $M_{V}$
values for red giants. For an M4 III with a B-V value $\sim$1.6, the observed
$M_{V}$ spans roughly from -2.0 to +1.5 \citep{Kov1998}. Using
measurements just for the roughly 49,000 stars with {\sl Hipparcos} parallax
measurement accuracy better than 20\%, one finds the range of $M_{V}$ is $-2.0$
to +1.0 \citep{ESA1997}. Adopting these values, we estimate the distance to
J205601, and hence to the Cygnus Loop, to be between 1.2 and 4.6 kpc.
If the $E(B-V)$ to J205601 is closer to $\sim$0.20 rather than 0.08 
due to extra extinction arising from the surrounding nebulosity, 
then this distance range decreases to 1.0 and 3.9 kpc.

%\citep{HG1975}   the Mv range at M4 III is -1.5 to +0.5
%\citep{Kov1998}  for a B-V = 1.6 (M4 III) the range of Mv is    -1.5 to +1.0 
%\citep{ESA1997} Hipparcos  for 49k stars with parallax measurements better than 20\% -1.5 to +1.0                                              -1.5
%                lower branch of the red giants
%\citep{Higgins2012}          for the majority of stars but with some faint extension down to +2.0

Distances to the Cygnus Loop much greater than $\sim$2 kpc are 
highly unlikely for several reasons. These include implied remnant size and age at
such distances given the remnant's observed X-ray and optical shock properties, along with 
filament proper motion relative to measured filament radial velocities.  

However, the fact that J205601 might well lie at distances much greater than 2
kpc raises the question of whether or not it actually lies, in fact, inside the Cygnus
Loop. After all, thousands of stars are projected within the remnant's
boundaries. Thus one can not, {\it{a piori}}, discount the possibility that this
red giant is simply a background star located far away and unrelated to both the Cygnus
Loop and the nebulosity described above seen in its direction.

While it is almost certain that the nebulosity seen toward J205601 represents
material shocked by the Cygnus Loop's shock front and thereby lies inside
the remnant, a direct physical connection of this red giant with the
nebulosity is not definitive. Our red spectrum of
J205601 revealed no H$\alpha$ emission that could strengthen the scenario of
on-going mass loss supporting the formation of the observed surrounding nebula.
There is also the requirement that J205601 be a relatively faint M4 red giant
(although not the faintest) in order to lie at a distance less than 2 kpc.

However, the combination of J205601 being a red giant, a type of star known to
undergo substantial mass loss, and it's location near the geometric centre of
a small bow-shaped nebula strongly suggests that J205601 lies physically
inside, and is the source of, the observed nebula.  The nebula's higher electron
densities compared to almost all of Cygnus Loop' filaments, along with its 
unusual bow-shaped morphology also point to an unusual origin. Furthermore, it is 
unlikely to be just a small, isolated ISM cloud that has been shocked by
passage of the remnant's shock front because it possesses quite a different
morphology from that of other known small ISM clouds overrun, shocked, and accelerated by the Cygnus
Loop's blast wave \citep{Fesen1992,Patnaude2002}.    

Thus the preponderance of evidence indicates that the red giant J205601 likely lies within the
Cygnus Loop and inside its surrounding nebula, therefore implying a distance to
the remnant on the high side of the 0.4 -- 2.0 kpc prior distance estimates
(Table 1). On the other hand, for J205601 to be inside the remnant, distances
less than 0.8 kpc are firmly excluded.  That is because even if J205601's luminosity
were that seen for the faintest observed M4 red giants 
(i.e., $M_{V} = +1.5$) and assuming an $E(B-V) = 0.2$, it's distance would be
$\sim$ 0.8 kpc. 

With an angular radius of $\sim$ 1 arcminute, the J205601 nebula 
is $\simeq0.3 \times$(d/1.0 kpc) pc in size.  Although some of its material could be red
giant mass loss material off J205601, a significant fraction is likely swept-up
ISM gas. Assuming an association with J205601 and an undecelerated red giant wind
velocity of 10 km s$^{-1}$, the nebula's dimension suggest a mass loss time
frame of $\lesssim$ 50,000 yr, with the nebula's double shell appearance suggesting
the mass loss occurred in two major episodes.

Distance estimates to the B star BD+31~4224 are more restrictive.  The most
quoted absolute Johnson $V$ magnitude for a B7 V star is $-0.40$
\citep{Jaschek1998,GC2009,mamajek2017}.  Adopting an apparent magnitude of 9.58
for BD+31~4224, a $M_{V}$ value of $-0.4$ and an $A_V = 0.25$ mag consistent
with its observed $E(B-V)$, translates to a distance of 880 pc.

However, a range of absolute Johnson magnitudes for B7 V stars have been cited
in recent literature, with \citet{mamajek2017} listing $M_{V}$ values from
$-0.11$ to $-0.67$; e.g., \citet{Lang1992} lists $M_{V} = -0.6$  while
\citet{Wegner2006} gives $-0.63$ based on {\sl Hipparcos} parallaxes
for 138 B7 V stars.  Adopting $M_{V}$ values of $-0.1$ to $-0.67$, estimated
distances to BD+31~4224 range from 770 to 1000 pc.

Since we were not able to completely rule out a B7 IV classification, we have
also calculated a maximum distance to BD+31~4224 if it were a B7 IV star. Using
$M_{V}$ values of $-0.77$ to $-1.3$ for B7 IV stars \citep{Wegner2006,GC2009},
we obtain distances of 1050 and 1330 pc.  However, given that BD+31~4224's
spectrum and colors match so well primary B7 V standards, we view smaller distances of
800 to 1000 pc as more likely. 

In summary, the B7 V star BD+31~4224 suggests a Cygnus Loop distance around  
0.8 to 1.0 kpc, while the M4 III star J205601 points to a distance of at
least 1 kpc. The fact that there is little overlap in the estimated distances
to two stars which we believe both lie inside the remnant, suggests that the
remnant lies at a distance of $\sim$1 kpc and hence farther than many
previous estimates. Based on distance estimates to these stars, 
but giving greater weight to smaller distance estimates
arising from the B7 star, we conclude the Cygnus Loop lies at a distance
of 0.8 - 1.2 kpc. 

% Figure 14
\begin{figure}
\begin{center}
\includegraphics[width=1.0\linewidth]{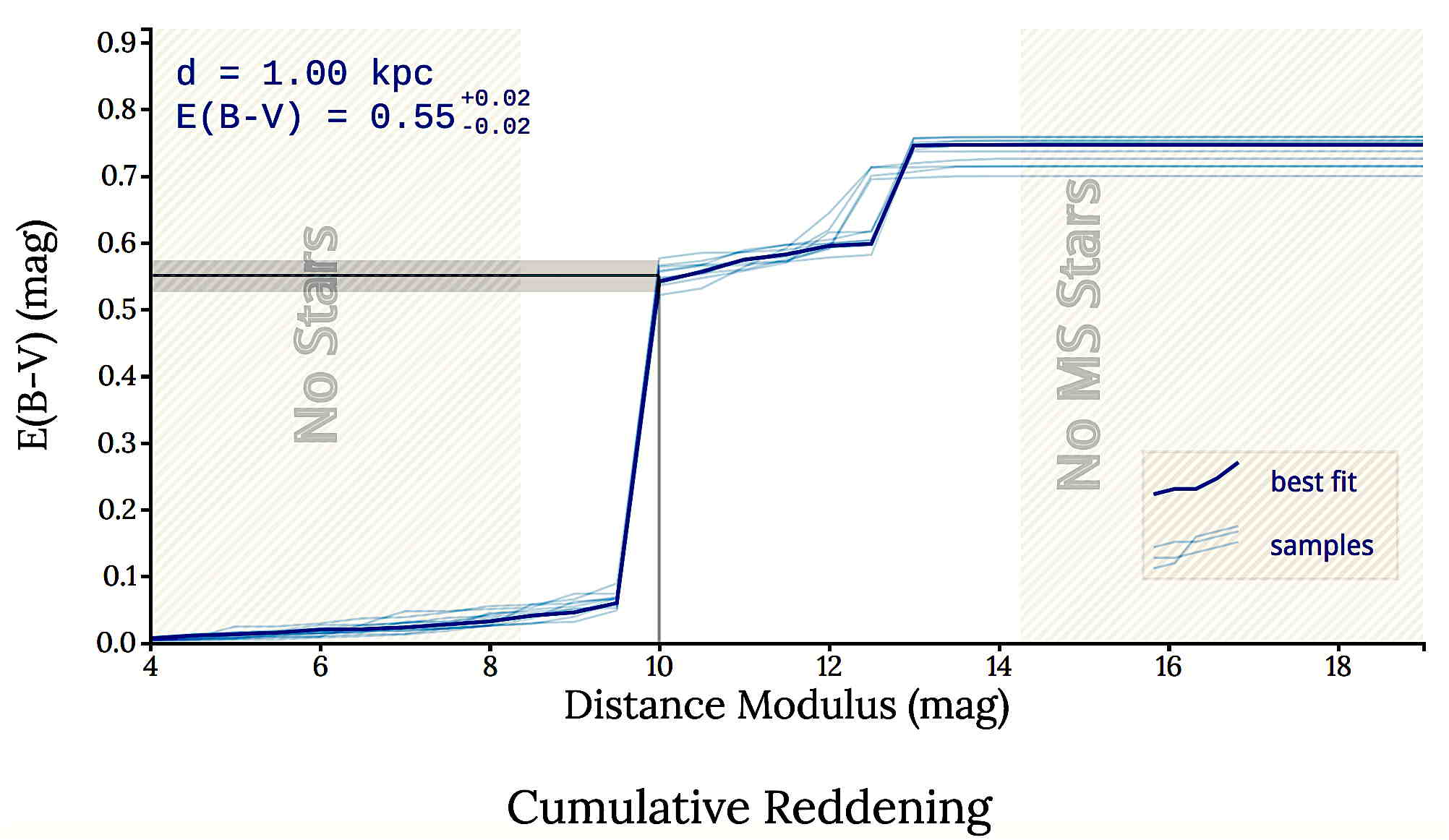}
\caption{Pan-STARRS1 derived plot of cumulative dust reddening 
along the western limb of the Cygnus Loop at $\alpha$[J2000] = $20^{\rm h}
45^{\rm m} 13^{\rm s}$, $\delta$[J2000] = $+30.5^{\rm o}$. Made using website: https://argonaut.skymaps.info. }
\end{center}
\label{dust_distance_plot}
\end{figure}

\subsection{A Distance Estimate Using Dust Reddening
along the Cygnus Loop's Western Limb}

We note that a distance to the Cygnus Loop $\simeq$ 1 kpc is consistent with
estimates for the distance to a large molecular cloud situated along with
remnant's western limb and long viewed as likely being impacted by the
remnant's shock front.  \citet{Duncan1923} described NGC~6960 as the ``frontier
between a region of many faint stars on the east and fewer on the west'', with
both \citet{Wolf1923} and \citet{Oort1946} noting that the remnant's bright
western nebulosity lies at the precise border of a dark nebula to the west.

The presence of dust from this molecular cloud is obvious on images published
by \citet{Ross1931} who noted that NGC 6960 lies on the boundary of a very
extensive dark nebula with striking differences in stellar density east and
west of the nebula.  Star density differences on either side of NGC 6960 is
most apparent in wide-angle blue images of the remnant and these differences
have been well studied by both \citet{Chamber1953} and \citet{Bok1957}. 

CO maps of \citet{Scoville1977} show the presence of a large molecular cloud
near the remnant's western emission (NGC~6960) with excellent correlation with
the observed optical obscuration.  \cite{Levenson1996} examined the optical and
X-ray emission along the western limb of the Cygnus Loop and concluded the
remnant was directly interacting with this cloud.

Figure 14 shows a plot of dust reddening with distance along the western limb
of the Cygnus Loop based on a three-dimensional map presented in
\citet{Green2015}, which utilizes Pan-STARRS \citep{Schlafly2014} and 2MASS
\citep{Skrutskie2006} photometry.  This figure shows a sharp rise of
extinction at 1 kpc, which then steadily increases from 1.0 out to 3.1 kpc,
then rises again out to 4 kpc where E(B-V) reaches 0.75 mag. The sharp rise of
extinction at 1 kpc is presumably due to the CO cloud the remnant appears to be
physically encountering and we take this as additional supporting evidence for
a distance to the Cygnus Loop of around 1 kpc.

The presence of some of the remnant's shock filaments some 25$'$ to 30$'$
farther to the west of the the Cygnus Loop's bright western nebula, NGC~6960, (see Fig.\
4 in \citealt{Fesen1992}) suggests the remnant lies close to the near side of
the molecular cloud. If correct, this then means that the remnant's distance
is probably closer to 0.8 -- 1.0 kpc than much larger values suggested by the M star. 

\section{Conclusions}

The Cygnus Loop is among the brightest and best studied Galactic supernova
remnants.  Unfortunately, like many other Galactic remnants, its distance
has not yet been determined to high accuracy. 

Here we present optical images and spectra of two small nebulosities with
projection locations within the Cygnus Loop supernova remnant which we suspect
are the results of stellar wind material interacting with the remnant's
expanding shock wave. We have identified one star within each of these two
nebulae which we propose as the source of their respective surrounding nebula
and use optical photometry and spectra to estimate their distances via
spectroscopic parallax. We then use the resulting stellar distance estimates to
then constrain the Cygnus Loop's distance. 

We find that an M4 III star located near the centre of a shocked, bow-shaped
nebula situated a few arcminutes west of the Cygnus Loop's bright eastern
nebula NGC~6992 lies at an estimated distance of between 1.0 and 4.6 kpc. 
A B7 V star located along the remnant's northwestern limb and centred in
an arc of shocked emission surrounded by a much larger region of curved and
twisted filaments likely lies at a distance of between 0.8 and 1.0 kpc. 

A Cygnus Loop distance of around 1 kpc would be consistent with the estimated
distance to a molecular cloud situated along the remnant's western limb with
which the remnant appears to be interacting. Thus, based on the assumption that
these two stars lie inside the remnant, combined with the estimated 
distance to a molecular cloud situated along the remnant's western limb, we
propose a distance to the Cygnus Loop of $1.0 \pm 0.2$ kpc.  A distance of 1.0
kpc implies a physical size for the remnant of $\approx 50 \times 60$ pc.

A distance around 1 kpc would help to resolve the issue of deduced postshock
cosmic ray to gas pressure ratios being near zero or negative for remnant
distances below 650 pc \citep{Salvesen2009}. Also a distance slightly greater
than 1 kpc would yield a SN explosion energy near the canonical SN explosion
energy of $10^{51}$ erg.  Such a distance could also explain the failure by
\citet{Welsh2002} to detect high-velocity \ion{Na}{1} and \ion{Ca}{2} line
absorptions associated with the remnant in several stars located in the
line-of-sight to the Cygnus Loop at distances up to 800 pc.     

If one or both of the two stars we have identified truly lie inside the Cygnus
Loop, then parallax measurements will finally provide us with an
accurate distance to the remnant.  The ESA parallax mission, {\sl GAIA}, which
can measure parallax values down to 24 micro-arcseconds, is capable of
providing this information.   

Finally, we note that \citet{Boubert2017}, who searched for runaway
former companions of the progenitors of nearby Galactic core-collapse supernova
remnants which included the Cygnus Loop, identified a 2 solar mass A type star
candidate runaway TYC 2688-1556-1.  However, they assumed a Cygnus Loop
distance of 0.54 kpc which, if our estimate of $\simeq$1 kpc is correct, means
this star is unlikely to be a runaway companion star associated with the Cygnus
Loop SN.

\acknowledgements

We thank Alex Brown and Tom Ayers for help in classifying the M star, John
Raymond and Bill Blair for valuable comments and discussions, Ignacio Cisneros
for help with examining the proper motions of the J205601 nebula's filaments, and the MDM
Observatory staff for their excellent instrument assistance.  This work was
made possible by funds from the NASA Space Grant, the Denis G. Sullivan Fund,
and Dartmouth's School of Graduate and Advance Studies.

\newpage

\end{document}